\input ./style/arxiv-general.cfg
\documentclass[aoas,MSNbibl,nameyear,dvips]{arximspdf}
\makeatletter
   \@ifpackageloaded{graphicx}{}{\usepackage{graphicx}}
\makeatother
\usepackage{multirow}
\usepackage{graphicx}

\doi{10.1214/15-AOAS805}% Updated by VTEXPTS2LaTeX.exe, 27.01.2015
\volume{9}
\issue{1}
\pubyear{2015}
\firstpage{474}
\lastpage{506}
\docsubty{FLA}

\makeatletter
\newcommand{\lleft}{\left}
\newcommand{\rright}{\right}
\newproclaim{exa}{Example}
\newtheorem{prop}{Proposition}
\newtheorem{lem}{Lemma}
\makeatother

\begin{document}
\begin{frontmatter}

\title{Network tomography for integer-valued traffic}
\runtitle{Network tomography for integer-valued traffic}

\begin{aug}
% Corresponding author: Martin Hazelton - m.hazelton@massey.ac.nz% Updated by VTEXPTS2LaTeX.exe, 27.01.2015 12:59
%Updated by VTEXPTS2LaTeX.exe, 27.01.2015 10:54
\author{\fnms{Martin L.}~\snm{Hazelton}\corref{}\ead
[label=e1]{m.hazelton@massey.ac.nz}}
\runauthor{M. L. Hazelton}
\affiliation{Massey University}
\address{Institute of Fundamental Sciences\\
Massey University\\
Palmerston North, 4442\\
New Zealand\\
\printead{e1}}
\end{aug}

% HISTORY:
%
\received{\smonth{3} \syear{2014}}% Updated by VTEXPTS2LaTeX.exe,
%27.01.2015 10:54
%
\revised{\smonth{1} \syear{2015}}% Updated by VTEXPTS2LaTeX.exe,
%27.01.2015 10:54

% ABSTRACT
%
\begin{abstract}
A classic network tomography problem is estimation of properties of the
distribution of
route traffic volumes based on counts taken on the network links. We
consider inference
for a general class of models for integer-valued traffic. Model
identifiability is examined.
We investigate both maximum likelihood and Bayesian methods of
estimation. In practice, these
must be implemented using stochastic EM and MCMC approaches. This requires
a methodology for sampling latent route flows conditional on the
observed link counts.
We show that existing algorithms for doing so can fail entirely,
because inflexibility in
the choice of sampling directions can leave the sampler trapped at a
vertex of the convex
polytope that describes the feasible set of route flows. We prove that
so long as the network's
link-path incidence matrix is totally unimodular, it is always possible
to select a coordinate
system representation of the polytope for which sampling parallel to
the axes is adequate. This
motivates a modified sampler in which the representation of the
polytope adapts to provide
good mixing behavior. This methodology is applied to three road traffic
data sets.
We conclude with a discussion of the ramifications of the unimodularity
requirements for the
routing matrix.
\end{abstract}

% KEYWORDS
\begin{keyword}
\kwd{EM algorithm}
\kwd{MCMC}
\kwd{origin--destination}
\kwd{polytope}
\kwd{transport}
\kwd{unimodular matrix}
\end{keyword}
\end{frontmatter}

%s1 #&#
\section{Introduction}

Network-based transport models occur commonly in road traffic
engineering and in the
analysis of electronic communications systems [e.g.,
\citet{CastroCoatesLiangNowakYu04,Denbyetal07}].
Such models can also be found in biological settings, for example, in
the study of
fungal networks [e.g., \citet{Heatonetal12}]. The implementation of
network models gives
rise to a variety of interesting and challenging statistical problems.
In particular, it
is frequently the case that observed data provide only indirect
information on many model
parameters of interest. We are then faced with network tomography, a
term coined by \citet{Vardi96}
to characterize the resulting types of inverse problems. This area has
received significant
attention over the past 20 years, with important contributions by
\citet{Vardi96,TebaldiWest98,CaoDavisVanderWielYu00,LiangYu03,LawrenceMichailidisNair06} and
\citet{AiroldiBlocker13}, among others.
\citet{CastroCoatesLiangNowakYu04} and \citet{Kolaczyk09} provide overviews.

A classical and heavily studied example of network tomography is the
problem of estimating
(mean) origin--destination (O--D) volumes from traffic counts at fixed
network locations, sometimes
referred to as volume network tomography. If~there
are multiple routes connecting some or all of the O--D pairs, then a
more general version of
the problem is to estimate the mean route flows. Inference would be
straightforward if we
observed the actual route flows, but the link count data determine them
only up to a system
of linear equations that is usually highly underdetermined.

Much of the early work on this problem appeared in the transportation
science literature, focusing on
road traffic networks. Underdetermination of the route flows was
addressed by either assuming
the existence of serviceable prior information [e.g., \citet{Maher83,Cascetta84,Bell91}] or by making very strong (and in some cases seemingly
arbitrary) modeling
assumptions [\citet{VanZuylenWillumsen80}]. One of the things that
characterized much
of this work was a focus on algorithms, with the underlying models
often poorly specified. In
particular, the distinction between the realized route flows and the
mean value thereof
was typically blurred.

\citet{Vardi96} introduced a more statistically principled approach to
the estimation
of mean O--D flows. Working under the assumption of Poisson distributed traffic,
he demonstrated identifiability of these parameters from sequences of
traffic count data and expounded on the difficulties of
likelihood-based inference. In particular,
he noted that the Poisson likelihood requires a sum over all feasible
route flow vectors: that is,
route flows that are consistent with the traffic counts and solve the
aforementioned linear system.
\citet{Vardi96} showed that this set will be far too large to enumerate
in anything other than toy
problems. We note that the inconvenient form of the likelihood is not
restricted to Poisson models.
The likelihood function for general integer-valued traffic models can
only be expressed as a
sum over the (typically huge) set of feasible route flows. See Section~\ref{seclik} for details.

The inferential problem can become far simpler if one is willing to
employ a continuous approximation
to discrete traffic flows. In that case the likelihood can be expressed
as an integral over all
feasible route flows. For normally distributed flows this integral can
typically be
evaluated analytically, a result that \citet{Vardi96} used to develop a
method-of-moments type
estimator for normal approximations to Poisson traffic models.
Normal models have formed the basis of the majority of work on O--D
matrix estimation (and similar
``passive'' network tomography problems) when applied to electronic
communication networks. See, for
example, \citet{CaoDavisVanderWielYu00,CastroCoatesLiangNowakYu04}. In
more complicated continuous
flow models the likelihood and/or Bayesian posterior may not be
available explicitly. However,
even then, working with continuous flows opens up MCMC sampling
approaches that are not available in
the discrete case. See, for example, \citet{AiroldiBlocker13} on
inference for their multilevel
state-space models of time series of (large) traffic flows in
communications networks.

For road traffic examples (which provide the author's motivating
interest), continuous flow
approximations are less attractive. In particular, even in large and
busy road networks there
will typically be large numbers of plausible routes with very small
(often zero) traffic counts.
With this in mind, we focus on models for integer-valued traffic flows.
A natural approach to
circumvent the impracticality of enumerating the feasible route flow
set for such models is
to sample therefrom. \citet{TebaldiWest98} were the first authors to
describe a comprehensive methodology of this type
when they studied Bayesian inference for Poisson models using an MCMC
algorithm that involved
conditional sampling of the latent route flows. The crucial step in
this work was the development
of a componentwise method for drawing feasible candidate route flows.
While this algorithm can
work adequately in benign examples, \citet{H10tech} and \citet
{AiroldiHass11} showed that it can mix
very slowly in more difficult problems, and even fail entirely in some
cases. Moreover, the
poor practical performance that is sometimes observed highlighted
critical gaps in our theoretical
understanding of the route flow sampling problem in general and errors
in the mathematical analysis
of the properties of the proposed sampler in particular. Despite the
pivotal importance of developing
a reliable route flow sampler for implementing likelihood-based methods
of inference, subsequent progress
on this sampling problem in the discrete flow case has been limited. In
particular, the only tangible
advance has been restricted to networks with particularly simple
topologies, like transit networks
and trees. See \citet{H10tech}.

In this paper we study network tomography for integer-valued traffic
flows for a rather general
class of models. We examine model identifiability [generalizing the
results of \citet{Vardi96}] and
consider both maximum likelihood and Bayesian inference implemented
through sampling-based methods.
We build on recent work by \citet{AiroldiHass11} and \citet
{AiroldiBlocker13} in the continuous flow case
to provide geometrical insight in the route flow sampling problem.
These methods help to provide
a better understanding of both the practical and theoretical aspects of
\citeauthor{TebaldiWest98}'s (\citeyear{TebaldiWest98})
sampler, and motivate a modified sampler with much improved properties.

Our work is largely focused on inference for static network parameters,
based on either a single
observed set of link counts or a sequence of such that can be regarded
as a random sample for
modeling purposes. Such problems are of significant interest in the
context of road network
planning, where, for example, static O--D matrices are frequently
employed when examining the
effects of proposed network changes. Nonetheless, the route flow
sampler that we develop could
equally well be employed within algorithms for fitting time-varying
models, such as those
developed by \citet{AiroldiBlocker13} for computer networks.

The remainder of the paper is organized as follows. We introduce our
general class of traffic models in
the next section and examine the foundations of inference for them in
Section~\ref{secinference}.
In Section~\ref{secsampling} we study the properties of \citeauthor
{TebaldiWest98}'s (\citeyear{TebaldiWest98}) route flow sampler
and introduce our modified version thereof. We illustrate the
application of our methodology
on traffic data from sections of the road network in Leicester in
Section~\ref{secapplication}. We draw
conclusions in Section~\ref{secdiscuss}, and discuss the practical and
theoretical consequences of
making convenient assumptions about the pattern of permissible routes
through the network

%s2 #&#
\section{Traffic models}

We consider a (weakly) connected network with $m$ nodes and $n_0$ links
(numbered sequentially
in both cases). Each traveler on the network makes a journey between an
origin and destination node.
Not all node pairs need be O--D pairs. We therefore introduce $\mathcal
{I}$ to
be the set of O--D node
pairs, with cardinality $c \le m(m-1)/2$. The elements of $\mathcal
{I}$ are
ordered lexicographically and so
may be referenced by a one-dimensional index.

Each O--D pair is connected by at least one route. In a big network with
reasonable connectivity there
will typically be a large number of possible routes. However, many of
these may be implausible in practice,
for example, because they contain cycles or are very circuitous. For
the sake of parsimony we choose to
ignore such routes in our model, and hence assume that a set of
permissible routes has been determined
{a priori}. We denote by $\mathcal{R}_k$ the set of (permissible) routes
for O--D pair $k \in\mathcal{I}$, and
let $\mathcal{R}= \bigcup_k \mathcal{R}_k$ be the set of all such
routes with
cardinality $r = |\mathcal{R}|$. We assume some
convenient ordering of the routes to allow one-dimensional indexing.

Let $\mathbf{x}= (x_1, \ldots, x_r)^{\mathsf{T}}$ denote the vector of
(integer-valued)
traffic flows on these routes
during some measurement period. These are not observed directly.
Instead our data comprise traffic counts
on $n \le n_0$ monitored links of the network. We denote these link
counts by $\mathbf{y}= (y_1, \ldots, y_n)^{\mathsf{T}}$.
They are related to the latent route flows by
%
%e1 #&#
\begin{equation}\label{eqfundamental}
\mathbf{y}= A \mathbf{x},
\end{equation}
where $A = (a_{ij})$ is the link-path incidence matrix defined by
\[
a_{ij} = \cases{
1,  & \quad\mbox{if link $i$ forms
part of route $j$},
\vspace*{3pt}\cr
0,  &\quad \mbox{otherwise}.}
\]
Typically $n \ll r$ so that (\ref{eqfundamental}) is a highly
underdetermined linear system. We denote
the set of nonnegative solutions of (\ref{eqfundamental}) (i.e.,  the
set of feasible route flows given
link counts $\mathbf{y}$) by $\mathcal{X}_{|\mathbf{y}} = \{ \mathbf
{x}\dvtx  \mathbf{y}= A \mathbf{x}, \mathbf{x}\ge
\mathbf{0} \}$, where the inequality is to
be interpreted elementwise.

For many problems it is natural to develop a statistical model of the
traffic system for which the parameters
relate directly to the route flows. We focus on such models and denote
by $f_X(\cdot| \bolds{\theta})$ the joint probability
mass function for $\mathbf{x}$. We assume that the support of $f_X$ is
$\mathcal{X}
= \mathbb{Z}^r_{\ge0}$.

\begin{exa}\label{exa1}
For estimation of mean O--D traffic flows
$\bolds{\theta}= (\theta_1, \ldots, \theta_c)$ in networks
with fixed routing (so that $r=c$ and route flows are identical to O--D
flows), \citet{Vardi96} and \citet{TebaldiWest98}
both assumed that $x_1, \ldots, x_r$ are independent with $x_j \sim
\operatorname{\mathsf{Pois}}(\theta_j)$. Traffic flows on real
road systems are often overdispersed in comparison to a Poisson model
[e.g., \citet{H01jrssc}], so use of
a negative binomial model is a plausible alternative.
\end{exa}

For a network with multiple routes per O--D pair, let $\mathbf{z}=
(z_1, \ldots
, z_c)^\mathsf{T}$ be the vector of O--D flows so
that $z_k = \sum_{j \in\mathcal{I}_k} x_j$. For estimation of
$\bolds{\mu}= \mathsf{E}[\mathbf{z}
]$ we could proceed
by aggregating results from a model where each route flow is separately
parameterized. For example, if we employ
the Poisson model $x_j \sim\operatorname{\mathsf{Pois}}(\theta_j)$
for independent route
flows $x_1, \ldots, x_r$, then $\mu_k = \sum_{j \in\mathcal{I}_k}
\theta_j$. We also obtain route choice probabilities as a by-product.
Specifically, the probability that a traveler
for O--D pair $k$ selects route $j \in\mathcal{I}_k$ is simply $p_j =
\theta
_j/\mu_{k(j)}$, where we use the notation $k(j)$
to emphasize that route $j$ connects O--D pair $k$.

An objection to this modeling approach is that it greatly exacerbates
the underdetermination of the fundamental
linear system (\ref{eqfundamental}), rendering statistical inference
all the more difficult. An alternative is to employ
a model for the route choice probabilities $\mathbf{p}= (p_1, \ldots,
p_r)^\mathsf{T}$ that is either partially or completely specified
exogenously. The Markov routing model examined by \citet{Vardi96} and
\citet{TebaldiWest98} does this in essence by
defining the route choice probabilities as the product of ``turning
probabilities'' at each node encountered en route.
However, the suitability of the underlying Markov assumption for real
road systems is highly questionable, not least
because it permits routes with (multiple) cycles. If we have travel
costs for the possible routes, then an alternative
is to employ random utility models [e.g., \citet{BenAkivaLerman85}].
Examples from the transport research literature
include various forms of logit route choice model [e.g., \citet
{DaganzoSheffi77,Cascettaetal96,KoppelmanWen00}]
and probit methods [e.g., \citet{Yaietal97}].

Even if a lightly parameterized route choice model is used, the number
of O--D pairs $c$ will typically exceed the number
of monitored links $n$ by a large margin. It follows that if we
observed just a single link count vector $\mathbf{y}$, then we
will require additional information in order to obtain unique point
estimates. In principle, more can be
learned from link count data $\{\mathbf{y}^{(t)}; t=1,2,\ldots,N \}$ collected
over a
sequence of observation periods. The subsequent analysis is most
straightforward if $\mathbf{y}^{(1)},\ldots,\mathbf{y}^{(N)}$ are
assumed to be independent and identically distributed. An alternative
is to model the inter-period dynamics of the
traffic flow as a Markov process [e.g., \citet{Cascetta89}]. However,
for most commonly used traffic models of this
type, the nature of the inferential problems remains essentially the
same [see \citet{ParryH13trb}].

%s3 #&#
\section{Tools for statistical inference}
\label{secinference}

%s3.1 #&#
\subsection{Model likelihood and identifiability}
\label{seclik}

Consider modeling a single link count vector $\mathbf{y}$. We can
derive the
likelihood function, $L$, by
conditioning on the latent trip vector:
%
%e2 #&#
\begin{eqnarray}
L(\bolds{\theta}) &=& f_Y(\mathbf{y}| \bolds{\theta})
\nonumber
\\[-8pt]
\label{eqlik}
\\[-8pt]
\nonumber
 &=& \sum_{\mathbf{x}} f_{Y|X}(\mathbf{y}|
\mathbf{x}, \bolds {\theta}) f_X(\mathbf{x}| \bolds{\theta})
= \sum_{\mathbf{x}\in\mathcal{X}_{|\mathbf{y}}} f_X(\mathbf{x}| \bolds{\theta}).
\end{eqnarray}
The third equality follows from the fact that $f_{Y|X}(\mathbf{y}|
\mathbf{x},
\bolds{\theta})$ is the indicator function
for the constraint $\mathbf{y}= A \mathbf{x}, \mathbf{x}\ge\mathbf
{0}$. Notice that this
is independent of $\bolds{\theta}$.
For general integer-valued traffic models it is not possible to
simplify (\ref{eqlik}). This means that
exact computation of the likelihood requires enumeration of the
feasible route flow set
$\mathcal{X}_{|\mathbf{y}} = \{ \mathbf{x}\dvtx  \mathbf{y}= A \mathbf
{x}, \mathbf{x}\ge\mathbf{0} \}$. For even
moderately sized networks this
will typically be computationally infeasible.

\begin{exa}\label{exa2}
Consider a sequence of 52 consecutively
numbered nodes connected in series,
with the first two being the origins and the last 50 being the
destinations of travel. Suppose
that 25 vehicles originate at each of the first two nodes, and that
each of the remaining 50 nodes is the
destination for a single vehicle, so that the link count vector is
given by $\mathbf{y}= (25,50,49,48,\ldots,2,1)$.
Despite the simplicity of the network and the presence of fixed
routing, there are nonetheless ${50 \choose 25}
> 10^{14}$ elements in the set $\mathcal{X}_{|\mathbf{y}}$.
\end{exa}

The likelihood from a random sample of $N$ traffic count vectors is
%
%e3 #&#
\begin{equation}\label{eqlik2}
L(\bolds{\theta}) = \prod_{t=1}^N \sum
_{\mathbf{x}^{(t)} \in
\mathcal{X}_{|\mathbf{y}^{(t)}}} f_X\bigl(\mathbf{x}^{(t)}
| \bolds{\theta}\bigr).
\end{equation}
Again, we cannot usually expect any simplification of his function.
This has ramifications for
the existence of nontrivial sufficient statistics. For example, suppose
that we employ the Poisson
model from Example~\ref{exa1}. If we were to observe the route flows directly,
then $\bar{\mathbf{x}} = N^{-1} \sum_{t=1}^N \mathbf{x}^{(t)}$
would be a sufficient statistic. However, with just link count data,
the only sufficient statistic
is the set of raw observations $\{\mathbf{y}^{(t)}; t=1,2,\ldots,N \}$. In
particular, the mean link count
vector $\bar{\mathbf{y}}$ is certainly not sufficient for $\bolds
{\theta}$.

It is intuitively obvious that the full sequence of link counts
contains more information than the mean
vector. In particular, the pattern of dependence between link counts is
illuminating in untangling
the indeterminacy problem. As an illustration based on Example~\ref{exa2}, note
that independence of $y_1$ (the
number of travelers leaving node 1) with all of $y_{27}, \ldots,
y_{51}$ (the numbers of travelers
arriving at nodes 28 to 52, resp.) occurs if and only if all
travelers originating at node 1
are destined for nodes numbered in the range $3$ to $27$. What is less
immediately apparent is
whether the information contained in the link counts is generally
sufficient to make the model parameters
identifiable.

\citet{Vardi96} addressed this issue in a particular case.
Specifically, he examined the identifiability
of $\bolds{\theta}$ from link count data in networks with fixed
routing when
the route flows are independent with
$x_j \sim\operatorname{\mathsf{Pois}}(\theta_j)$ for $j=1, \ldots,
r$. He showed that if the
columns of $A$ are distinct and each
has at least one nonzero entry, then $\bolds{\theta}$ is identifiable.
Subsequent work on identifiability has
focused primarily on models for computer networks, where the large
traffic counts justify the use of
continuous flow distributions with support that is not explicitly
bounded at zero. The most comprehensive
contribution is due to \citet{SinghalMichailidis07}. They derived
results that can be applied to models with
certain types of spatio-temporal dependence between route flows, but
placed restrictions on the traffic routing
schemes and network structure.

We seek to extend \citeauthor{Vardi96}'s (\citeyear{Vardi96}) result to general routing schemes and
discrete traffic distributions
while maintaining the generality of permissible network structures (in
terms of topology and placement
of traffic counters). As the following theorem shows, the assumptions
of Poisson route flows and fixed
routing can be relaxed while maintaining identifiability of the model
parameters.

\begin{prop}\label{prop1}
Let the columns of $A$ be distinct, with each
containing at least one nonzero element.
Assume that the route flows are independent and that the marginal
distribution of each has support
equal to the nonnegative integers. Then if the model parameter vector
is identifiable from independent
observations on $\mathbf{x}$, it is also identifiable from independent
observations on $\mathbf{y}$.
\end{prop}

The proof is given in the \hyperref[app]{Appendix}.

In principle, this is a reassuring result, indicating that if we
observe an independent sequence of link
counts over our network, then we can eventually hope to obtain unique
parameter estimates despite the underlying
structural ambiguities. However, the nature of our constructive proof
suggests that it could require an
excessively long sequence of observations on $\mathbf{y}$ to do so. In
practice, we may have limited ability to untangle
the elements of $\bolds{\theta}$ from even quite lengthy sequences of
link counts.
%, even if we are able to assume
%this sequence forms a random sample, an unrealistic hope in many
%applications.

%s3.2 #&#
\subsection{Sampling-based inference}
\label{secsampling-inference}

Exact inference based on the model likelihood is typically not possible
because enumeration of the
set $\mathcal{X}_{|\mathbf{y}}$ is computationally infeasible. A
natural alternative
is to use an approximation
to the likelihood where the summation over all elements of $\mathcal
{X}_{|\mathbf{y}
}$ in (\ref{eqlik}) is replaced
by a sum over some suitable sample therefrom. The essence of this idea
can be implemented using the
stochastic EM algorithm for purely likelihood-based inference, and via
MCMC methods in a Bayesian setting.

We consider first the EM algorithm when a random sample $\mathbf{y}
^{(1)},\ldots,\mathbf{y}^{(N)}$ of link counts is
available. Regarding the route flows as missing data, the complete data
likelihood is
\[
\prod_{t=1}^N f_{X,Y}\bigl(
\mathbf{y}^{(t)}, \mathbf{x}^{(t)} | \bolds {\theta}\bigr) = \prod
_{t=1}^N f_X\bigl(
\mathbf{x}^{(t)} | \bolds{\theta}\bigr)
\]
since $\mathbf{y}^{(t)}$ is a deterministic function of $\mathbf
{x}^{(t)}$. The
expectation of the complete data
log-likelihood computed with respect to the distribution of the route
flows conditional on the
link counts is given by
%
%e4 #&#
\begin{equation}\label{eqEM}
Q\bigl(\bolds{\theta}| \bolds{\theta}'\bigr) = \sum
_{t=1}^N \sum_{\mathbf
{x}^{(t)} \in\mathcal{X}_{|\mathbf{y}
^{(t)}}} \log
\bigl\{ f_X\bigl(\mathbf{x}^{(t)} | \bolds{\theta}\bigr)
\bigr\} f_{X|Y}\bigl(\mathbf{x}^{(t)} | \mathbf{y}
^{(t)}, \bolds{\theta}'\bigr)
\end{equation}
for any given $\bolds{\theta}'$. The algorithm\vspace*{1pt} proceeds by iterating between
finding the maximizer $\tilde{\bolds{\theta}}$
of $Q$ and computation of $Q(\bolds{\theta}| \bolds{\theta}')$ with
$\bolds{\theta}'$ reset
to equal $\tilde{\bolds{\theta}}$. It
converges to the maximum likelihood estimate $\hat{\bolds{\theta}}$.

Evaluation of the conditional expectation $Q(\bolds{\theta}| \bolds
{\theta}')$ is
difficult because of the
summation over feasible route flow sets in (\ref{eqEM}), although for
Poisson models the problem can be
simplified in certain special cases as noted by \citet
{VanderbeiIannone94} and \citet{Li05}. As an alternative,
one could consider using a normal approximation if the flows are not
too small [e.g., \citet{Vardi96,Li05}].
A more generally applicable approach\vspace*{1.5pt} is to approximate $Q(\bolds
{\theta}|
\bolds{\theta}')$ by computing
the mean of each term $\log\{ f_X(\mathbf{x}^{(t)} | \bolds{\theta
}) \}$ over $M$
simulations drawn from $f_{X|Y}(\mathbf{x}^{(t)} | \mathbf{y}^{(t)},
\bolds{\theta}')$. Accordingly, the stochastic EM algorithm works by replacing
$Q(\bolds{\theta}| \bolds{\theta}')$ by
%
%e5 #&#
\begin{equation}\label{eqMCEM}
\hat{Q}\bigl(\bolds{\theta}| \bolds{\theta}'\bigr) =
M^{-1} \sum_{i=1}^M \sum
_{t=1}^N \log\bigl\{ f_X\bigl(
\mathbf{x}^{*(t)}_i | \bolds{\theta}\bigr) \bigr\},
\end{equation}
where $\mathbf{x}^{*(t)}_1, \ldots, \mathbf{x}^{*(t)}_M$ is a random
sample from
$f_{X|Y}(\cdot| \mathbf{y}^{(t)}, \bolds{\theta}')$. Importantly,
when implementing the stochastic EM algorithm in practice, the number
of simulations $M$ should adapt as the
algorithm progresses in order to ensure convergence. See \citet
{CaffoJankJones05}. This is illustrated
in the application studied in Section~\ref{secapp1}.

Standard errors for $\hat{\bolds{\theta}}$ can be obtained via the
missing information
principle in the usual way [e.g., \citet{Louis82,Tanner96}]. The
observed information matrix is estimated by
%
%e6 #&#
\begin{eqnarray}
I_{\mathrm{obs}} &\equiv& I_{\mathrm{obs}}\bigl(\hat{\bolds{\theta}};
\mathbf{y}^{(1)}, \ldots, \mathbf{y}^{(N)}\bigr)
\nonumber
\\
\label{eqMCEMse}
&=& \frac{1}{M} \sum_{i=1}^M
\bigl(I\bigl(\hat{\bolds{\theta}}; \mathbf{x}^{*(1)}_i,
\ldots, \mathbf{x}^{*(N)}_i\bigr)
\\
&&\hspace*{10pt}\qquad{}-\mathbf{u}\bigl(\hat{\bolds{\theta}}; \mathbf {x}^{*(1)}_i,
\ldots, \mathbf{x} ^{*(N)}_i\bigr) \mathbf{u}\bigl(\hat{
\bolds{\theta}}; \mathbf{x}^{*(1)}_i, \ldots,
\mathbf{x}^{*(N)}_i\bigr)^\mathsf{T} \bigr),\nonumber
\end{eqnarray}
where\vspace*{1pt} $\mathbf{u}({\bolds{\theta}}; \mathbf{x}^{*(1)}_i, \ldots,
\mathbf{x}^{*(N)}_i) = \sum_{t=1}^N \partial\log\{ f_X(\mathbf{x}^{*(t)}_i | \bolds{\theta
})\}/\partial\bolds{\theta}$
is the complete data score vector and $I({\bolds{\theta}}; \mathbf
{x}^{*(1)}_i, \ldots, \mathbf{x}^{*(N)}_i) = - \partial\mathbf{u}({\bolds{\theta}};
\mathbf{x}^{*(1)}_i, \ldots, \mathbf{x}^{*(N)}_i)
/\partial\bolds{\theta}^\mathsf{T}$ the complete data information
matrix. The
(approximate) variance--covariance
matrix of $\hat{\bolds{\theta}}$ is given by $I_{\mathrm{obs}}^{-1}$.

Turning to Bayesian inference, suppose that we have available a prior
$\pi(\bolds{\theta})$ for the model
parameters. Exact computation of the posterior $p(\bolds{\theta}|
\mathbf{y})
\propto\pi(\bolds{\theta}) L(\bolds{\theta})$ will
generally be infeasible because of the difficulties in computing the
likelihood. Computation of the
normalizing constant for $p(\bolds{\theta}| \mathbf{y})$ is an
additional problem.
We therefore resort to MCMC
methods to generate posterior samples.

Following the lead of \citet{TebaldiWest98}, we avoid the need to
enumerate all feasible route flows by
sampling from the joint posterior of $\bolds{\theta}$ and $\mathbf
{x}^{(1)}, \ldots,
\mathbf{x}^{(N)}$. Working in the case
of origin--destination matrix estimation with $N=1$, Tebaldi and West (\citeyear{TebaldiWest98}) proposed a Gibbs sampler,
iterating between draws of $\bolds{\theta}$ and $\mathbf{x}$ from
their respective
conditional distributions. The
former conditional simplifies: $p(\bolds{\theta}| \mathbf{x},
\mathbf{y}) = p(\bolds{\theta}| \mathbf{x}
)$ because $\mathbf{y}$ is determined
by $\mathbf{x}$. It follows that conditional sampling of $\bolds
{\theta}$ will
typically be straightforward. For instance,
for the Poisson traffic model in Example~\ref{exa1} we get a gamma conditional
for $\bolds{\theta}$ when using conjugate
gamma priors for the components of $\bolds{\theta}$. We may employ
Metropolis--Hastings sampling for nonconjugate
models. At the second stage of each iteration, conditional sampling of
$\mathbf{x}$ requires draws from
$f_{X|Y}(\cdot| \mathbf{y}, \bolds{\theta})$.

Applying the same methodology to alternatively parameterized traffic
models and to sample sizes $N > 1$
presents no further problems in principle, although there is
flexibility as to the order in which variables
are updated. For example, one can choose to intersperse updates of
$\bolds{\theta}$ between draws of each of
$\mathbf{x}^{(1)}, \ldots, \mathbf{x}^{(N)}$. Indeed, there is considerable
flexibility in the overall design of the
sampler that may lead to improved convergence properties. For instance,
in the context of origin--destination
matrix estimation, \citet{AiroldiBlocker13} recommend modifying
\citeauthor{TebaldiWest98}'s (\citeyear{TebaldiWest98}) scheme to
use a Metropolis--Hastings algorithm where candidate pairs $(\theta_i,
x_i)$ are sampled.

%s4 #&#
\section{Conditional sampling of route flows}
\label{secsampling}

%s4.1 #&#
\subsection{Overview of the problem}
\label{secsampling-overview}

The methods of sampling-based inference discussed in the previous
section all share a common problem: the
need to sample efficiently the latent route flow vector $\mathbf{x}$ given
observed link counts $\mathbf{y}$. The development
of an effective method of doing so has proved challenging. The only
published methodology for integer-valued
flows on general networks is due to \citet{TebaldiWest98}. More
recently, \citet{AiroldiHass11} and \citet{AiroldiBlocker13}
made progress on the continuous flow version of the problem. In this
section we review these methods and give
examples where \citeauthor{TebaldiWest98}'s (\citeyear{TebaldiWest98}) sampler fails. By
examining the geometry of the feasible set $\mathcal{X}_{|\mathbf{y}}$
we are able to explain this behavior and also prove results
characterizing the conditions under which convergence
is guaranteed. This in turn leads us to propose a modified version of
\citeauthor{TebaldiWest98}'s (\citeyear{TebaldiWest98}) methodology with
far better properties.

We frame the problem in terms of a Metropolis--Hastings sampler for the
conditional distribution $f_{X|Y}(\cdot| \mathbf{y}, \bolds{\theta})$,
the current state of which is $\mathbf{x}\in\mathcal{X}_{|\mathbf
{y}}$. The general
approach is to generate a candidate
vector $\mathbf{x}^\dagger$ from a proposal distribution $q$ with support
$\mathcal{X}_{|\mathbf{y}}$ which is then accepted with probability
$\min(\alpha,1)$, where
%
%e7 #&#
\begin{eqnarray}
\alpha&=& \frac{f_{X|Y}(\mathbf{x}^\dagger| \mathbf{y}, \bolds
{\theta}) q(\mathbf{x}
)}{f_{X|Y}(\mathbf{x}| \mathbf{y}, \bolds{\theta})q(\mathbf
{x}^\dagger)}
\nonumber\\
 \label{eqalpha}
&=& \frac{f_{X,Y}(\mathbf{x}^\dagger, \mathbf{y}| \bolds{\theta})
q(\mathbf{x})}{f_{X,Y}(\mathbf{x}, \mathbf{y}
| \bolds{\theta})q(\mathbf{x}^\dagger)}
\\
\nonumber
&=& \frac{f_{X}(\mathbf{x}^\dagger| \bolds{\theta}) q(\mathbf
{x})}{f_{X}(\mathbf{x}| \bolds{\theta})
q(\mathbf{x}^\dagger)}.
\end{eqnarray}
Note that the final equality in (\ref{eqalpha}) holds only if
$\mathbf{x}^\dagger$ is feasible, which will always be the case
if $q$ has support $\mathcal{X}_{|\mathbf{y}}$. It is possible to
employ $q$ with
support that extends beyond $\mathcal{X}_{|\mathbf{y}}$ on
the understanding that any infeasible vectors $\mathbf{x}^\dagger$ are
automatically rejected. Nonetheless, this will be a practical
option only if the effective support of $q$ is a reasonable
approximation to $\mathcal{X}_{|\mathbf{y}}$, otherwise the
acceptance rate is
liable to be unacceptably low.

Such a sampler can be initialized at a feasible route flow vector in a
number of ways. For example, we may use standard
integer programming methods to obtain the optimal element of $\mathcal{X}
_{|\mathbf{y}}$ against some prespecified criterion.

%s4.2 #&#
\subsection{Tebaldi and West's sampler}

Assume that the rows of $A$ are linearly independent. (If this is not
the case, then it indicates that one or more
of the link counts is redundant and can be omitted from the analysis
without any loss of information.) Then the
observed link counts place $n$ linear constraints on the route flow
vector~$\mathbf{x}$. As \citet{TebaldiWest98} note,
if we reorder the routes (and hence columns of~$A$) in a suitable
manner, then we can write (\ref{eqfundamental}) as
%
%e8 #&#
\begin{equation}\label{eqdecompose1}
\mathbf{y}= [ A_1 | A_2 ] \lleft[
\matrix{ \mathbf{x}_1
\vspace*{3pt}\cr
\mathbf{x}_2}
 \rright] = A_1
\mathbf{x}_1 + A_2 \mathbf{x}_2,
\end{equation}
where $A_1$ is an $n \times n$ invertible matrix. It follows that
%
%e9 #&#
\begin{equation}\label{eqdecompose2}
\mathbf{x}_1 = A_1^{-1}(\mathbf{y}-
A_2 \mathbf{x}_2),
\end{equation}
indicating that we need sample only elements of the ($r-n$)-dimensional
vector $\mathbf{x}_2$.

The conditional distribution of $\mathbf{x}_2$ given $\mathbf{y}$ has support
\[
\mathcal{X}_{2|\mathbf{y}} = \{ \mathbf{x}_2\dvtx  \mathbf{y}= A \mathbf
{x}, \mathbf{x}\ge\mathbf{0} \} = \bigl\{\mathbf{x}_2 \dvtx
A_1^{-1}(\mathbf{y}- A_2
\mathbf{x}_2) \ge\mathbf{0}, \mathbf {x}_2 \ge\mathbf{0}
\bigr\},
\]
where the vector inequalities are to be interpreted elementwise.
Attempting to sample $\mathbf{x}_2$
{en bloc} would require a convenient characterization of $\mathcal{X}
_{2|\mathbf{y}}$, which we do not
have. \citet{TebaldiWest98} therefore suggested componentwise sampling
of $\mathbf{x}_2$. In describing this technique
we employ the route ordering implied in (\ref{eqdecompose1}) so that
$\mathbf{x}_1 = (x_1, \ldots, x_n)^\mathsf{T}$
and $x_{n+j}$ is the $j$th element of $\mathbf{x}_2$ for $j=1,\ldots,(r-n)$.
We let $\mathbf{x}_{2,-j}$ denote the
vector $\mathbf{x}_2$ with its $j$th element omitted.

Let\vspace*{-2pt} us then consider updating $\mathbf{x}_{2}$ by sampling $x^\dagger_{n+j}$
conditional on $\mathbf{x}_{2,-j}$
and~$\mathbf{y}$. As \citet{TebaldiWest98} showed, the conditional
support of
$x^\dagger_{n+j}$ is a finite sequence of
contiguous integers. They did not compute the endpoints of this
sequence explicitly, relying instead on an
acceptance--rejection methodology to \mbox{generate} a feasible candidate.
However, computation of the
endpoints is straightforward, as we discuss later. We may therefore
sample $x^\dagger_{n+j}$ from a
distribution~$q_{2,j}$ with appropriate support. Defining $\mathbf
{x}^\dagger
_{2}$ to be $\mathbf{x}_{2}$ with just
the $j$th component updated in this manner, we then compute $\mathbf
{x}^\dagger
_1$ according to (\ref{eqdecompose2})
to give the full candidate vector of route flows $\mathbf{x}^\dagger$. This is
accepted with probability $\min(\alpha_j,1)$,
where
%
%e10 #&#
\begin{equation}\label{eqalphaj}
\alpha_j = \frac{f_{X|Y}(\mathbf{x}^\dagger| \mathbf{y}, \bolds
{\theta})
q_j(x_{n+j})}{f_{X|Y}(\mathbf{x}| \mathbf{y}, \bolds{\theta
})q_j(x^\dagger_{n+j})} = \frac{f_{X}(\mathbf{x}^\dagger| \bolds{\theta})
q_j(x_{n+j})}{f_{X}(\mathbf{x}| \bolds{\theta}
) q_j(x^\dagger_{n+j})}.
\end{equation}
For models with {a priori} independent route flows, the
contributions to $f_X$ on the top and bottom
will cancel for all routes except $1, 2, \ldots, n, n+j$.

Sequentially sampling the elements $\{x_{n+j}, j=1, \ldots, n-r \}$ in
this manner will guarantee eventual
sampling from $f_{X|Y}(\cdot| \mathbf{y}, \bolds{\theta})$ so long
as $\mathcal{X}_{2|\mathbf{y}
}$ is connected, in the sense that it is
possible to move between any two elements of this space by a sequence
of moves parallel to the coordinate
axes. Unfortunately this is not always the case.

\begin{exa}\label{exa3}
Figure~\ref{fig1} depicts a shortened version of the series network
examined in Example~\ref{exa2}. We assume that
nodes 1 and 2 are the only origins of flow, and nodes 3, 4, 5 are the
only travel destinations. If we order
the routes lexicographically by origin then destination, the
(partitioned) link-route incidence matrix is
given by
%
%e11 #&#
\begin{equation}\label{eqegA}
A = [A_1 | A_2 ] = \lleft[
\begin{array}{c@{\quad}c@{\quad}c@{\quad}c@{\hspace*{6pt}}|@{\hspace*{6pt}}c@{\quad}c} 1 & 1 & 1 & 0 & 0 & 0
\\
1 & 1 & 1 & 1 & 1 & 1
\\
0 & 1 & 1 & 0 & 1 & 1
\\
0 & 0 & 1 & 0 & 0 & 1 \end{array}
 \rright].
\end{equation}
Note that the matrix $A_1$ in this partition is invertible, as required
for application of
\citeauthor{TebaldiWest98}'s (\citeyear{TebaldiWest98}) sampler.

%f1
%f1 #&#
\begin{figure}

\includegraphics{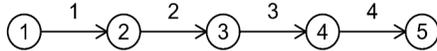}

\caption{An example four link network.}\label{fig1}
\end{figure}

Suppose that we observe link counts $\mathbf{y}= (10, 20, 20,
10)^\mathsf{T}$. Given
that node 3 is not a source of
travel, we can immediately infer that the route flows destined for this
node are both zero, that is,
$x_1 = x_4 = 0$. It is then straightforward to show that the feasible
route set is defined by $\mathcal{X}_{2|\mathbf{y}}
= \{ x_5 + x_6 = 10\}$, as displayed in the left-hand panel of Figure~\ref{fig2}. In this situation,
\citeauthor{TebaldiWest98}'s (\citeyear{TebaldiWest98}) sampler
will fail entirely, since there
are no feasible steps in the coordinate
directions of $x_5$ and $x_6$.

Such an extreme situation could be avoided by pre-checking whether any
route flows are uniquely determined
by the observed link counts. However, even if we discount such cases,
it is simple to construct examples
where the sampler mixes extremely slowly. For instance, suppose that
$\mathbf{y}= (10, 20, 19, 9)^\mathsf{T}$, implying
that a single traveler is destined for node 3. The corresponding set of
feasible route flows is defined
by $\mathcal{X}_{2|\mathbf{y}}$ as displayed in the right-hand panel
of Figure~\ref{fig2}. Generation of candidates in
the coordinate directions of $x_5$ and $x_6$ will allow route flows to
change by no more than one unit,
so that exploration of the entirety of $\mathcal{X}_{|\mathbf{y}}$
will be a
laborious business. Moreover, we can
design cases where mixing becomes arbitrarily slow by increasing flows
for all routes except those
destined for node 3. For instance, if $\mathbf{y}= (1000, 2000, 1999,
999)^\mathsf{T}
$, then the feasible set will
appear as a much thinner version of the right-hand panel of Figure~\ref{fig2}.

%f2
%f2 #&#
\begin{figure}

\includegraphics{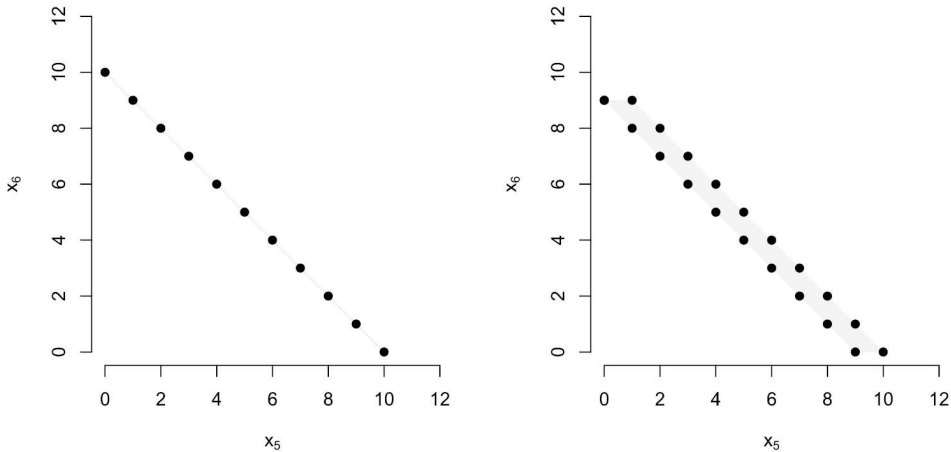}

\caption{Feasible route sets in terms of flows for routes
5 (from node 2 to node 4) and 6 (from node 2 to node 5) for the
network in Figure~\protect\ref{fig1}. The left-hand panel corresponds to the
case $\mathbf{y}= (10, 20, 20, 10)^\mathsf{T}$; the right-hand
panel to $\mathbf{y}= (10, 20, 19, 9)^\mathsf{T}$. The interior of
the convex hull of
each feasible set is shaded as an embellishment
to help emphasize shape.}\label{fig2}
\end{figure}

Working in the context of continuous traffic flow modeling, \citet{AiroldiBlocker13} addressed
these problems by selecting a random search direction in $\mathcal
{X}_{2|\mathbf{y}
}$. A~candidate $\mathbf{x}$ is
then sampled from along the correspondingly oriented line segment.
However, adapting this approach
to integer-valued flows is not straightforward, since appropriate
discretization of the sampling
probabilities requires computationally expensive information on the
local geometry of $\mathcal{X}_{|\mathbf{y}}$.
Furthermore, this random directions algorithm will mix poorly in cases
like Example~\ref{exa3} because ``long
moves'' will be achieved only infrequently when a fortuitous
orientation is proposed.
\end{exa}

%s4.3 #&#
\subsection{A modified route flow sampler}
\label{secnewalg}

Intuitively, we can think about the route flows corresponding to the
columns of $A_1$ as providing a
``swap space'' in \citeauthor{TebaldiWest98}'s (\citeyear
{TebaldiWest98}) algorithm, in the
sense that if we wish to transfer travelers
between routes, then this can only occur by swapping the travelers in
and out of this space.
When seeking to update $x_{n+j}$ (for $j \ge1$),
the value of this flow can only increase or decrease to the extent that
we can obtain travelers from,
or donate travelers to, the route flows in the swap space. Looking back
at Example~\ref{exa3}, the problem is
that $x_1$ and $x_4$ can take only a very small range of values. It is
the resultant lack of slack in
the swap space that prevents the sampler from taking large steps and
mixing well.

This discussion motivates a simple modification to \citeauthor{TebaldiWest98}'s (\citeyear{TebaldiWest98}) algorithm. The decomposition
of $A$ will typically be far from unique, even given that $A_1$ must be
invertible. We should then seek to
reorder the routes, and hence adjust the partition of $A$, so as to
increase the slack available in the
swap space. We can demonstrate formally that this will work, in the
sense that it is always possible, and
moreover practicable, to find a partition of $A$ with sufficient slack
to allow the sampler to mix adequately.
Our developments make use of some results in integer geometry, building
on the work of \citet{AiroldiHass11}.

In what follows, we make the assumption that the matrix $A$ is totally
unimodular. That is, each invertible
square submatrix (and hence any $A_1$ that we consider) is integer
valued. The requirement that $A_1^{-1}$
be an integer matrix is implicit in the work of \citet{TebaldiWest98},
since otherwise there is no certainty
that all the sampled route flows will be integers. This assumption is
explicitly stated in the
theoretical work of \citet{AiroldiHass11} and \citet{AiroldiBlocker13}.
As the last mentioned authors
note, total unimodularity appears to be a very common property of
link-route incidence matrices; it holds for
all examples that they found in the literature. Nonetheless, it is not
assured, a matter that we discuss in more
detail in Section~\ref{secdiscuss}.

Consider any partition $A = [A_1 | A_2]$ for which $A_1$ is invertible,
and define
%
%e12 #&#
\begin{equation}\label{eqU}
U = \lleft[ \matrix{
-A_1^{-1}
A_2
\vspace*{3pt}\cr
I_{r-n} }
 \rright],
\end{equation}
where $I_{r-n}$ is the ($r-n$)-dimensional identity matrix. As \citet{AiroldiHass11} note,
$AU = 0$, so that the columns of $U$ generate the null space of $A$.
Moreover, because $A$
is totally unimodular, $U$ is integer valued. Now, the $j$th column
$\mathbf{u}
_j$ of $U$ (for
$j=1,\ldots,r-n$) is zero in all components below the $n$th, apart from
$u_{n+j,j} = 1$. It follows
that if $\mathbf{x}\in\mathcal{X}_{|\mathbf{y}}$, then $\mathbf
{x}\mp\mathbf{u}_j$ is potentially the
vector of flows
resulting from transferring a single traveler to or from the swap space
to route $n+j$.
We say potentially because there is no certainty that $\mathbf{x}\mp
\mathbf{u}_j$
will be feasible.

\citeauthor{TebaldiWest98}'s (\citeyear{TebaldiWest98}) algorithm
works by iteratively sampling
in the directions
$\mathbf{u}_1, \dots, \mathbf{u}_{r-n}$. For a given partition of
$A$ this will work
so long as
movement from $\mathbf{x}\in\mathcal{X}_{|\mathbf{y}}$ is possible
in at least one of
those directions.
If movement is possible parallel to $\mathbf{u}_j$, then, as noted
before, the
(conditional)
support of a candidate $x^\dagger_{n+j}$ is a contiguous integer sequence
$\{\chi_{\mathrm{lo}}, \chi_{\mathrm{lo}}+1, \ldots, \chi_{\mathrm{hi}} \}$, where the endpoints
are given by
\[
\chi_{\mathrm{lo}} = \max \bigl( - \max\bigl\{-x^*_j \dvtx
u_j = 1 \bigr\}, 0 \bigr)
\]
and
\[
\chi_{\mathrm{hi}} = \max \bigl( \min\bigl\{x^*_j\dvtx
u_j = -1 \bigr\}, 0 \bigr),
\]
in which $(x^*_1, \ldots, x^*_n)^\mathsf{T}= \mathbf{x}^* = A_1^{-1}
(\mathbf{y}- A_{2,-j}
\mathbf{x}_{2,-j})$ with $A_{2,-j}$
being the matrix $A_2$ with the $j$th column deleted.

Nevertheless, as we saw in Example~\ref{exa3}, there is no guarantee that
movement is possible in
any of the aforementioned directions. We note that this contradicts
the irreducibility result given in the Appendix of \citet{TebaldiWest98}, but the proof
therein is flawed since it relies on the fallacious premise that if all
the elements of
a sum of vectors are nonnegative, then at least one of the summands
must have no negative
elements. What we will show is that when the algorithm is stuck at
$\mathbf{x}
$ for a given
ordering of the columns of $A$, there is always an alternative
partition $[A_1 | A_2]$
such that movement is possible parallel to a column of the adjusted~$U$.

Let us temporarily relax the requirement that the route flows be
integer valued. Then
geometrically, the feasible set $\mathcal{X}_{|\mathbf{y}}$ for
real-valued flows is
formed by the
intersection of the linear manifold $\{\mathbf{x}\dvtx \mathbf{x}= A
\mathbf{y}\}$ with the
nonnegative
orthant $\{\mathbf{x}\dvtx \mathbf{x}\ge\mathbf{0} \}$. The
resulting set forms a convex
polytope [see \citet{Ziegler95}, e.g.]. This is an ($r-n$)-dimensional object embedded within $r$
dimensional space. It can be characterized by the convex hull of its
vertices, where $\mathbf{x}\in\mathcal{X}_{|\mathbf{y}}$
is a vertex of the polytope if it has $r-n$ zero coordinates.
Furthermore, we have the following:

\begin{lem}[{[\citet{AiroldiHass11}]}]\label{lem1}
The vertices of $\mathcal
{X}_{|\mathbf{y}}$
are integer valued even when the route
flows are continuous.
\end{lem}

We note in passing that while this result is stated and proved in \citet{AiroldiHass11}, it has been
known for some time. Equivalent results can be found in \citet{HoffmanKruskal56} and \citet{VeinottDantzig68}.

Suppose $\mathbf{x}$ is a vertex. We can reorder the columns of $A$ so that
the last $r-n$ entries of $\mathbf{x}$ are
zero. (The matrix $A_1$ under this reordering must be invertible,
otherwise $\mathbf{x}$ would be infeasible and
hence not a vertex.) Therefore, $\mathbf{x}+ \mathbf{u}_j$ (for any
$j=1, \ldots,
r-n$) has $r-n-1$ zero elements,
is integer valued and, if feasible, lies on an edge of the polytope.
Now, for any
general point $\mathbf{x}\in\mathcal{X}_{|\mathbf{y}}$ that is not
a vertex, convexity of
the polytope ensures that movement
must be possible parallel to some edge. In order to prove that the
sampler will mix, it remains to show that
the sampler cannot get stuck at a vertex. There are two possibilities.
If movement is not possible along
any edge leading from $\mathbf{x}$, then this vertex must be the sole element
of $\mathcal{X}_{|\mathbf{y}}$, in which
case the route flows are uniquely determined by the link counts. If
movement is possible parallel to
the $j$th column of $U$, then $\mathbf{x}+ \mathbf{u}_j \in\mathcal
{X}_{|\mathbf{y}}$. That we
can take an integer-valued
step in that direction is assured by Lemma~\ref{lem1}.

We have proved the following:

\begin{prop}\label{prop2}
Given any feasible integer-valued flow vector $\mathbf{x}$, either
\begin{longlist}[(ii)]
\item[(i)] $\mathbf{x}$ is the sole element of $\mathcal
{X}_{|\mathbf{y}}$; or
\item[(ii)] there exists a matrix partition $A = [A_1, A_2]$ and
corresponding matrix $U$ from (\ref{eqU})
such that $\mathbf{x}+ \mathbf{u}_j \in\mathcal{X}_{|\mathbf{y}}$
for some $1 \le j \le r-n$.
\end{longlist}
\end{prop}

This result ensures that the sampler will always have a feasible
integer-valued move in at least one coordinate
direction, but provides no guidance as to whether movement is possible
in any given direction. The
next proposition provides a sufficient condition that tallies with our
earlier intuition about the
need for slack in the swap space. Specifically, if there is flow on all
the routes corresponding
to the first $n$ columns of $A$, then the sample has feasible moves
parallel to all the coordinate
axes defined by $U$.

\begin{prop}\label{prop3}
Let $\mathbf{x}\in\mathcal{X}_{|\mathbf{y}}$ with $x_i > 0$ for
$i=1,\ldots,n$. Then for
each $j=1,\ldots,r-n$, $\mathbf{x}+ \mathbf{u}_j$
is a feasible route flow vector.
\end{prop}

The proof is given in the \hyperref[app]{Appendix}.

\begin{figure}

\includegraphics{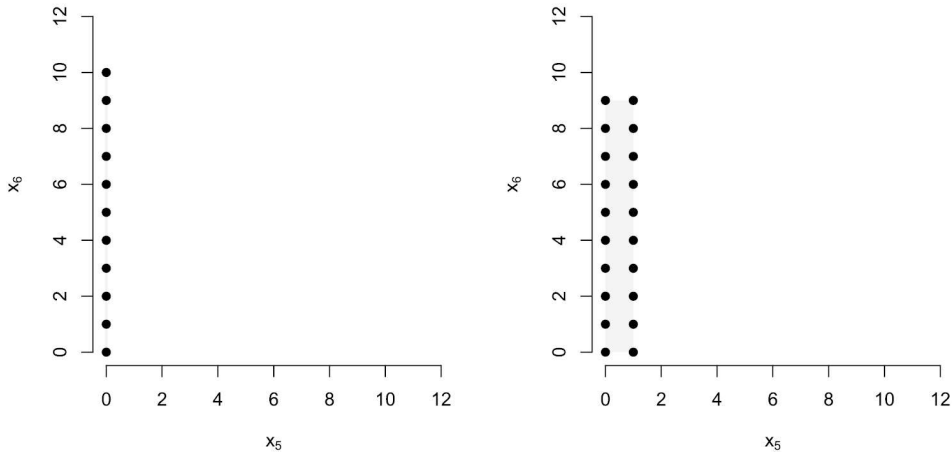}

\caption{Feasible route sets in terms of flows for
revised routes~5 (from node 2 to node 3) and 6 (from node 2 to node 5) for
the network in Figure~\protect\ref{fig1}. The left-hand panel corresponds to
the case $\mathbf{y}= (10, 20, 20, 10)^\mathsf{T}$; the~right-hand
panel to $\mathbf{y}= (10, 20, 19, 9)^\mathsf{T}$. The interior of
the convex hull of
each feasible set is shaded as an embellishment
to help emphasize shape.}\label{fig3}
\end{figure}

Altering the partition of $A$ corresponds to a change in the ($r-n$)-\break dimensional coordinate system
representation of $\mathcal{X}_{|\mathbf{y}}$. In particular, we can
choose a
representation in which one of the
axes is parallel to any given edge. This immediately explains how we
can hope to avoid the difficulties
encountered in Example~\ref{exa3}. The problem in Figure~\ref{fig2} is the
orientation of the polytope, rather
than the fact that it is long and thin. Let us switch columns 4 and 5
of $A$ to give
%
%e13 #&#
\begin{equation}\label{eqegA2}
A = [A_1 | A_2 ] = \lleft[ %
\begin{array}
{c@{\quad}c@{\quad}c@{\quad}c@{\hspace*{6pt}}|@{\hspace*{6pt}}c@{\quad}c} 1 & 1 & 1 & 0 & 0 & 0
\\
1 & 1 & 1 & 1 & 1 & 1
\\
0 & 1 & 1 & 1 & 0 & 1
\\
0 & 0 & 1 & 0 & 0 & 1 \end{array} %
 \rright].
\end{equation}
The polytope $\mathcal{X}_{|\mathbf{y}}$ is now represented in terms
of the route
flows from node 2 to node 3 (column 5)
and from node 2 to node 5 (column 6). The resulting feasible regions in
this coordinate system are displayed
in Figure~\ref{fig3} for the cases $\mathbf{y}= (10, 20, 20,
10)^\mathsf{T}$ and
$\mathbf{y}= (10, 20, 19, 9)^\mathsf{T}$. Clearly, sampling parallel
to the
coordinate axes will be efficient.

%f3
%f3 #&#

The preceding theory implies that convergence of the sampler can be
guaranteed if we update the partition of $A$ from
iteration to iteration in a suitable manner. In particular, a
sufficient condition for irreducibility will be that every
possible partition $A=[A_1 | A_2]$ (with $A_1$ invertible) is employed
infinitely often in the long run. This could be
achieved by systematically cycling through all possible partitions or
by sampling the partition at each iteration. Such
sampling would need to place nonzero probability on each partition, but
would work best if there is a bias toward selecting
partitions in which the first $n$ routes tend to carry high flows.

A direct implementation of this methodology would be feasible in
examples with modest numbers of routes. However,
in large examples the need to repeatedly find acceptable partitions of
$A$ would be computationally impractical.
Nonetheless, in such cases Proposition~\ref{prop3} will generally come to the
rescue, since it implies that we need only find
a single good partition that can then be used unchanged thereafter.
Specifically, if we can find $n$ linearly
independent columns of $A$ for which the corresponding route flows have
negligible posterior weight at (and preferably
near to) zero, then the sampler will work adequately if these columns
are selected to form $A_1$.

With these comments in mind, we recommend a phased approach. We start
with some initial partition of $A$ and run
the sampler for a fixed number of iterations. At the end of this first
pilot phase we compute pilot estimates from
the sampled route flows and use these to determine a suitable
permutation of the columns of $A$. In the next pilot phase we
run the sampler for another fixed number of iterations, refine our
estimates of the route flows and update the
partition of $A$ accordingly. This process can be continued until we
have discovered a suitable route ordering,
although in practice we have found that two pilot phases are typically
sufficient. Once the pilot phases are
completed, the sampler can run with no further changes to $A$.

During this process, each update of the partition of $A$ should be
chosen so that the routes corresponding to
the columns of $A_1$ carry relatively high flow. There are two issues
to consider. First, what statistic
should we compute as a summary of the magnitude of the sampled flows
for these routes? An obvious answer (and the
one that we employ in later examples)
is to use the mean flows from the pilot samples. Employing an estimate
of a very low percentile is an alternative
that relates directly to the desire to avoid very small flows on these
routes. The second issue concerns optimization
of the partition of $A$ based on the pilot estimates. In principle, we
could search through all possible partitions
of $A$ to find the one for which the sum of the pilot estimates is
largest, subject to the requirement that $A_1$
is invertible. A cheap alternative (used in subsequent numerical work)
is to employ a greedy sequential algorithm,
where the set of columns of $A_1$ is built up one route at a time, at
each step choosing the highest flow route that
is not in the span of the columns already selected.

Using this cheap version of the phased approach results in an algorithm
with \mbox{almost} exactly the same computational
expense per iteration as the original methodology of \citet{TebaldiWest98}. The additional computing time required
to generate a few additional matrix partitions is negligible. Of
course, we expect our algorithm to have far better
mixing properties in many applications, and when this is so, it will be
far cheaper in terms of the computational cost
per effective independent sample.

A comparison with the computational cost of \citeauthor{AiroldiBlocker13}'s (\citeyear{AiroldiBlocker13}) algorithm is a
little more involved.
This algorithm shares the same computational complexity as that of
\citet{TebaldiWest98} and the cheap (phased) version
of our refinement thereof, in the sense that the problem of generating
a new candidate route flow is fundamentally $O(r-n)$
in all cases. However, we expect the actual computing time per
candidate route flow
for \citeauthor{AiroldiBlocker13}'s (\citeyear{AiroldiBlocker13})
algorithm to be more than twice
that of ours because of the extra calculations
required to sample along random search directions. Nonetheless, it is
important to recognize that a proposed update of all components
of the route flow vector (which we refer to as a single iteration in
the numerical studies in the following section) requires
generation of $(r-n)$ candidates using our algorithm (one for each
column of $A_2$), while a single candidate from
\citeauthor{AiroldiBlocker13}'s (\citeyear{AiroldiBlocker13})
algorithm can update all components
of $\mathbf{x}$ simultaneously. The issue of which is more computationally
efficient in practice will depend upon acceptance rates. We consider
this matter a little further in the numerical
example in Section~\ref{seclargenet}, although it should be kept in
mind that our algorithm and that of \citet{AiroldiBlocker13}
are only approximately comparable, in the sense that they are designed
for discrete and continuous flows, respectively.

%s5 #&#
\section{Applications}
\label{secapplication}

Traffic models of the type that we have studied are used in practice to
model networks of various
scales, ranging from inter-urban motorway systems to individual urban
road intersections. In this
section we start by considering two applications of the latter type,
which provide convenient examples
to assess and illustrate our methods. We then go on to examine a larger
section of road network. The
first of the applications includes link count data from multiple days,
but no prior information, so
estimates are computed using maximum likelihood estimation through the
stochastic EM algorithm. In
the other examples we have data only from a single day, but informative
priors are available, making
Bayesian inference (via the MCMC algorithm) natural. All the
applications are taken from the road
system in the English city of Leicester.

We consider three algorithms for route flow sampling at various points
during this section. These
are \citeauthor{TebaldiWest98}'s (\citeyear{TebaldiWest98})
algorithm, our modification thereof
(with at most two partition
updates for $A$), and\break \citeauthor{AiroldiBlocker13}'s (\citeyear
{AiroldiBlocker13}) algorithm with
output rounded to integer
values. Additional numerical results detailing trace plots, effective
sample size and mean slack for
these samplers are available as part of the supplementary material for
this article [\citet{H15aoas-supp}].

%s5.1 #&#
\subsection{Application 1: Maximum likelihood
estimation for flows at an intersection}
\label{secapp1}

\begin{figure}[t]

\includegraphics{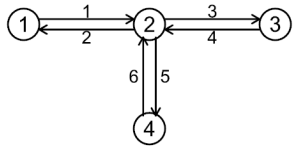}

\caption{An abstraction of the intersection of London
Road (running left to right) and University
Road in the English city of Leicester.}\label{fig4}
\vspace*{-6pt}
\end{figure}

\begin{figure}
\vspace*{-6pt}

\includegraphics{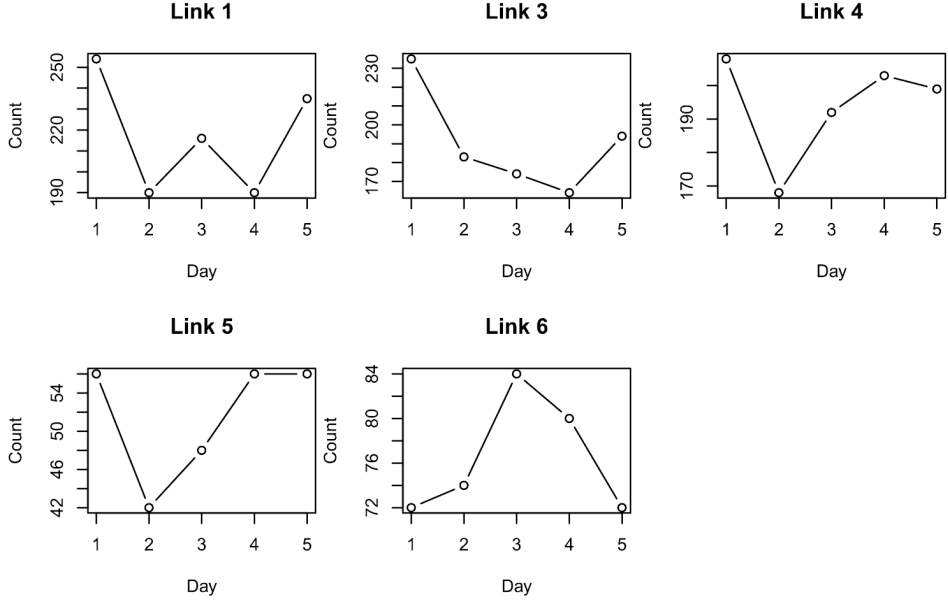}

\caption{Traffic counts for five (nonconsecutive) days
for the network in Figure~\protect\ref{fig4}.}\label{fig5}
\end{figure}

The first application concerns the intersection of London Road with
University Road in Leicester. An
abstracted form of the physical network is displayed in Figure~\ref{fig4}. All nodes are both origins
and destinations of travel, except for node 2 which is neither. All
links except for 2 are equipped with
inductive loop counters. We have available traffic flows on all other
links for the period 16:00--16:15 on
five nonconsecutive weekdays in May. The data are displayed as line
plots in Figure~\ref{fig5}.

%f4
%f4 #&#
%f5
%f5 #&#

We consider the problem of estimating the mean origin--destination
flows. With only 6 routes [one for each
of the O--D pairs $\{(i,j); i,j=1,3,4, i \ne j\}$] and 5 monitored
links, the linear system (\ref{eqfundamental})
is only slightly under-determined. Moreover, the ordering of the
columns of $A$ is irrelevant, since the feasible
route flow polytope is one-dimensional. The interest lies in whether a
sample of size $N=5$ is sufficient to allow
useful inferences to be made about the mean route flows based on the
likelihood alone (i.e., in the absence of prior
information) and to what extent model misspecification may effect the results.

We consider two models. The first is the Poisson model introduced in
Example~\ref{exa1}, while the second
is a negative binomial model, parameterized so that $\mathsf{E}[x_j] =
\theta
_j$ and $\operatorname{\mathsf{Var}}[x_j] =(1 + \alpha) \theta_j$
for $j=1,\ldots,6$. For both models we compute maximum likelihood
estimates using the stochastic
EM algorithm. The initial simulation size was set to $M=2000$ (following
a burn-in period of 2000 iterations). We then followed the strategy of
\citet{CaffoJankJones05} to control increases
in simulation size and provide the stopping rule. Standard errors were
computed using (\ref{eqMCEMse}).
For comparison, we also calculate parameter
estimates using \citeauthor{Vardi96}'s (\citeyear{Vardi96}) method of moments approach
applied only to the Poisson model. (This methodology cannot be employed
for the negative binomial model
because of the presence of the additional dispersion parameter.) The
results are displayed in Table~\ref{tabres1}.

%t1 #&#
\begin{table}[b]
\tabcolsep=0pt
\tablewidth=\textwidth
\caption{Estimates of mean route flows for Application
1. Maximum likelihood estimates (MLEs) were
calculated using the stochastic EM algorithm. MoM denotes
\citeauthor{Vardi96}'s (\citeyear{Vardi96}) method of moment estimator. For the
negative binomial model, the maximum likelihood estimate (and standard
error) of the dispersion parameter $\alpha$
was 1.92 (0.87)}
\label{tabres1}
\begin{tabular*}{\tablewidth}{@{\extracolsep{\fill}}lcccc@{}}
\hline
& & \multicolumn{2}{c}{\textbf{Poisson} \textbf{model}} & \\[-4pt]
&& \multicolumn{2}{l}{\hrulefill} \\
\textbf{Route} & \textbf{O--D} & \textbf{MLE} \textbf{(std err)} & \textbf{MoM} &\multicolumn{1}{c@{}}{\multirow{2}{95pt}[11pt]{\centering\textbf{Negative
binomial model  MLE (std err)}}} \\
\hline
1 & 1--3 & 175.5 (7.0) & 420.2 & 176.4 (12.9) \\
2 & 1--4 & \phantom{0}41.5 (4.7) & \phantom{0}37.5 & \phantom{0}41.0 (9.7)\phantom{0} \\
3 & 3--1 & 183.9 (7.1) & 155.4 & 183.5 (13.1) \\
4 & 3--4 & \phantom{0}10.1 (3.9) & \phantom{0}39.8 & \phantom{0}10.6 (8.1)\phantom{0} \\
5 & 4--1 & \phantom{0}61.9 (5.1) & \phantom{0}49.7 & \phantom{0}63.1 (10.4) \\
6 & 4--3 & \phantom{0}14.5 (4.0) & \phantom{00}0.0 & \phantom{0}12.8 (8.2)\phantom{0} \\
\hline
\end{tabular*}
\end{table}

The raw link counts suggest overdispersion with respect to a Poisson
model, and this is borne out
by the estimate of $\hat\alpha= 1.92$ ($\mathit{SE}=0.87$) for the dispersion
parameter in the negative
binomial model. Nonetheless, the maximum likelihood estimates of $\hat
{\bolds{\theta}}$ are very similar
for the two models.

We note that there is a marked difference in the nominal standard
errors obtained
between the estimates from the Poisson and negative binomial models. In
part this reflects the
relative capabilities of the models to account for the aforementioned
overdispersion. However, we also
note that inaccuracy in the standard errors can be expected given that
they rely on asymptotic likelihood
theory being applied to data from just $N=5$ time points.

To examine the properties of maximum likelihood estimation in this
application, we simulated 100 sets of
route flows from a negative binomial model with parameters matching
those estimated from the real data.
We then computed the corresponding link count data sets and found the
maximum likelihood estimates
for each using the stochastic EM algorithm using both a (correctly
specified) negative binomial model
and an (incorrectly specified) Poisson model. From these results we
calculated the (approximate) biases
of the estimates. These were low in all cases. The estimated bias in
$\hat\theta_i$ was less than $1\%$ for routes
$i=1, 3, 5$, which carry the heaviest traffic, rising to a maximum
(absolute) value of $3.7\%$ for route
$i=4$, which carries the lightest flow. The differences in bias between
the negative binomial and Poisson
models were negligible.

These results suggest not only that maximum likelihood can provide
useful estimates in this application, but
also that the estimates are quite robust to misspecification between
negative binomial and Poisson models.
Loosely speaking, maximum likelihood estimation based on the Poisson
model privileges first order information
over higher order information, in part evidenced by the fact that
$A\hat{\bolds{\theta}}_{\mathrm{pois}} = \bar{\mathbf{y}}$, where $\hat{\bolds{\theta}}_{\mathrm{pois}}$
denotes\vspace*{1pt} the vector Poisson maximum likelihood estimates and
$\bar{\mathbf{y}}= \mathsf{N}^{-1} \sum_{t=1}^N \mathbf{y}^{(t)}$.
The same comment does not apply to \citeauthor{Vardi96}'s (\citeyear{Vardi96}) method of moments
approach based\vspace*{1pt} on a Poisson model, which
produces highly implausible estimates $\tilde{\bolds{\theta}}_{\mathrm{mom}}$. For the
results of the real data analysis\vspace*{1pt}
reported in Table~\ref{tabres1}, the method of moments estimated
vector of mean link counts
$A\tilde{\bolds{\theta}}_{\mathrm{mom}}$ differs from $\bar{\mathbf{y}}$ by
more than a factor
of two in some components.

%s5.2 #&#
\subsection{Application 2: Bayesian inference at an intersection}

\begin{figure}

\includegraphics{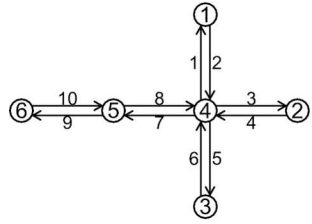}

\caption{An abstraction of the network around the
intersection of Regent Road (running left to right)
and University Road.}\label{fig6}
\end{figure}

The second network that we consider describes an area around the
junction of University Road and Regent Road.
See Figure~\ref{fig6}. All nodes are both origins and destinations of
travel, except for node 4 which is neither.
All links except for 5 are equipped with inductive loop vehicle
detectors. In this case we have available only
a single set of traffic counts, $\mathbf{y}=
(72,56,217,120,119,127,178,117,181)^\mathsf{T}$,
again collected on a weekday in May. We aim to estimate the mean route
flows $\bolds{\theta}$ via
the Poisson model from Example~\ref{exa1}. Prior information is essential in
this case and is available
from an outdated traffic survey. We incorporate this in the form of
pseudo route traffic counts
$\breve{\mathbf{x}}$ through independent Gamma priors with $\theta_j
\sim
\operatorname{\mathsf{Gamma}}(\breve x_j/2, 1/2)$.

%f6
%f6 #&#

We initialized the MCMC algorithm from the previous section with the
column order in $A$
randomized, subject to the requirement the $A_1$ be invertible in the
usual partition. This
gave
%
%e14 #&#
\begin{eqnarray}
\qquad A &=& [A_1 | A_2 ]
\nonumber
\\[-8pt]
\label{eqAnetB}
\\[-8pt]
\nonumber
 &=& \lleft[ %
\begin{array}{c@{\hspace*{9pt}}c@{\hspace*{9pt}}c@{\hspace*{9pt}}c@{\hspace*{9pt}}c@{\hspace*{9pt}}c@{\hspace*{9pt}}c@{\hspace*{9pt}}c@{\hspace*{9pt}}c@{\hspace*{5pt}}|@{\hspace*{5pt}}c@{\hspace*{9pt}}c@{\hspace*{9pt}}c@{\hspace*{9pt}}c@{\hspace*{9pt}}c@{\hspace*{9pt}}c@{\hspace*{9pt}}c@{\hspace*{9pt}}c@{\hspace*{9pt}}c@{\hspace*{9pt}}c@{\hspace*{9pt}}c} 0 & 0 & 0 & 1 & 1 & 0 & 0 & 0 & 0 & 0 & 0 & 0 & 0 & 0 & 0
& 1 & 0 & 0 & 0 & 1
\\
0 & 1 & 0 & 0 & 0 & 0 & 0 & 1 & 1 & 0 & 0 & 0 & 0 & 1 & 0 & 0 & 0 & 0 & 0 & 0
\\
0 & 0 & 1 & 0 & 0 & 0 & 1 & 0 & 0 & 0 & 0 & 0 & 1 & 1 & 0 & 0 & 0 & 0 & 0 & 0
\\
0 & 0 & 0 & 0 & 0 & 1 & 0 & 0 & 0 & 1 & 1 & 0 & 0 & 0 & 0 & 1 & 0 & 0 & 0 & 0
\\
1 & 0 & 0 & 0 & 1 & 0 & 0 & 0 & 0 & 0 & 0 & 1 & 1 & 0 & 0 & 0 & 0 & 0 & 0 & 0
\\
1 & 0 & 0 & 0 & 0 & 1 & 0 & 1 & 1 & 1 & 0 & 1 & 0 & 0 & 0 & 0 & 0 & 0 & 0 & 0
\\
0 & 0 & 1 & 1 & 0 & 0 & 1 & 0 & 0 & 0 & 0 & 0 & 0 & 0 & 0 & 0 & 0 & 1 & 1 & 1
\\
1 & 0 & 0 & 0 & 0 & 1 & 0 & 0 & 1 & 0 & 0 & 0 & 0 & 0 & 1 & 0 & 0 & 0 & 0 & 0
\\
0 & 0 & 0 & 1 & 0 & 0 & 1 & 0 & 0 & 0 & 0 & 0 & 0 & 0 & 0 & 0 & 1 & 0 & 1 & 0
\end{array}
 \rright].\hspace*{-6pt}
\end{eqnarray}
The initial flow pattern $\mathbf{x}$ was generated through integer
programming as the solution to
(\ref{eqfundamental}) with maximum $L_1$ norm.
%f7 #&#
\begin{figure}

\includegraphics{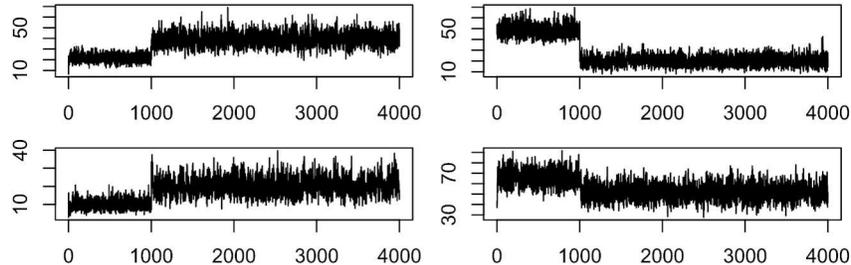}

\caption{Trace plots for mean route flows $\{\theta_j,
j=1,2,9,13\}$. Routes
are numbered according to the columns of $A$ in (\protect\ref{eqAnetB}), with
the matrix of plots filled
by row. Trace plots for all 20 routes are available as supplementary material.}\label{fig7}
\end{figure}

%f8 #&#
\begin{figure}[b]

\includegraphics{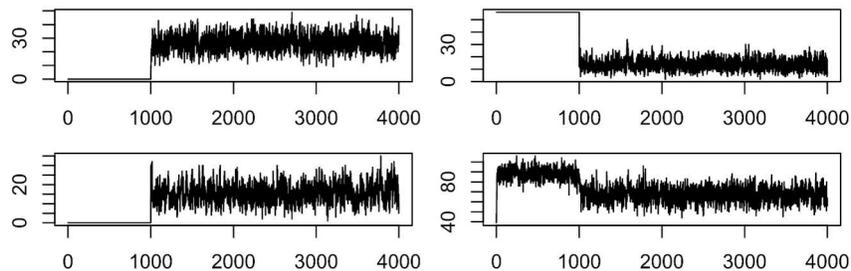}

\caption{Trace plots for sampled route flows $\{x_j,
j=1,2,9,13\}$. Routes
are numbered according to the columns of $A$ in (\protect\ref{eqAnetB}), with
the matrix of plots filled
by row. Trace plots for all 20 routes are available as supplementary material.}\label{fig8}
\end{figure}

The two pilot phases of the MCMC algorithm ran for $10{,}000$ iterations
each. At the end of the second phase
the partition of $A$ was updated so that the final column permutation was
$\sigma= (17,7,13,1,20,6,11,10,9,2,14,12,19,\break 16,8,18,3,15, 4,5)$ in
comparison to the initial ordering.
The algorithm then ran for a further burn-in period of $10{,}000$,
followed by a further $10{,}000$ iterations
from which posterior estimates were computed. The computing time for
this complete simulation was $37$ seconds,
with the algorithm coded in R [\citet{citeR13}] running on a 32-bit
Windows desktop computer with a dual core 3.6~GHz
processor and 4~GB of memory. Trace plots for all iterations for $\theta
_j$ and $x_j$ appear in Figures~\ref{fig7}
and \ref{fig8} for an illustrative selection of routes, specifically,
those numbered $j=1,2,9,13$, based on ordering
of the columns in (\ref{eqAnetB}). We see that while the algorithm is
not entirely stuck during the first
pilot phase, nonetheless only some route flows are successfully
updated. It follows that the unmodified version
of \citeauthor{TebaldiWest98}'s (\citeyear{TebaldiWest98}) sampler
would fail to converge to
the correct posterior distribution in this application.

The posterior means with corresponding 95\% credible intervals appear
in Table~\ref{tabres2} alongside
the prior values. Comparing the prior and posterior means, the overall
level of traffic is very
similar, but there are some marked differences in the pattern of flows.

%s5.3 #&#
\subsection{Application 3: Inference for a larger network}\label{seclargenet}

\begin{figure}

\includegraphics{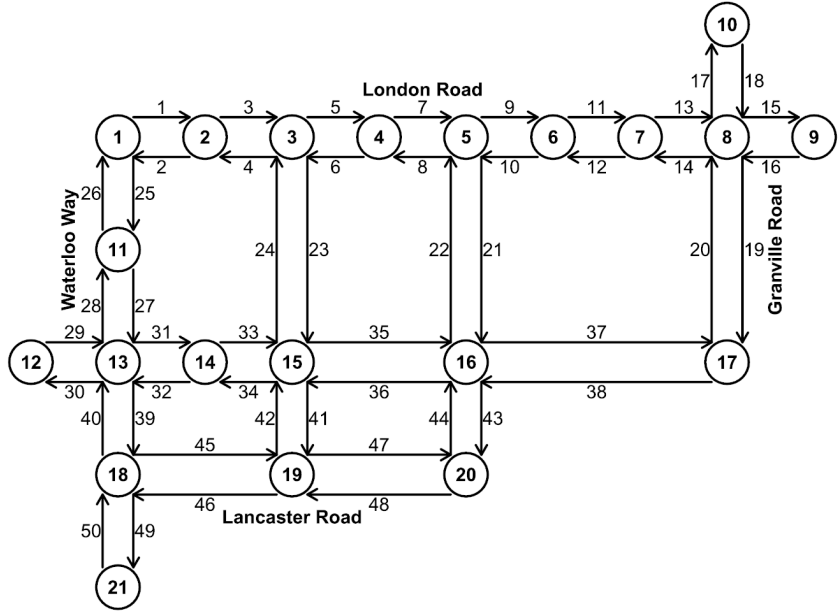}

\caption{A section of the road network in Leicester, just
to the southeast of the city
center.}\label{fig9}
\end{figure}

We now turn to a larger application based on a section of the city road
network just to the southeast of
the center of Leicester. See Figure~\ref{fig9}. This network has 21
nodes and 50 links. A total of 85
O--D pairs with an aggregate of 127 routes were considered, based on
earlier analyses of this network
[e.g., \citet{H01jrssc}]. A single set of traffic counts $\mathbf{y}$ is
available on 27 of the network links.
See Table~\ref{tab3} for details. As before, prior information is
essential and is available in the form of pseudo route
traffic counts $\breve{\mathbf{x}}$ based on an outdated survey.
These are
used to define independent Gamma priors
with $\theta_j \sim\operatorname{\mathsf{Gamma}}(\breve x_j/5,
1/5)$. We note that this type
of O--D matrix updating problem
is very common in transport planning and modeling.

%t2 #&#
\begin{table}
\caption{Prior and posterior means, 95\% posterior
credible intervals, for mean
route flows for Application~2 using a Poisson model}
\label{tabres2}
\begin{tabular*}{\tablewidth}{@{\extracolsep{\fill}}lcccc@{}}
\hline
\textbf{Route} & \textbf{O--D} & \textbf{Prior mean} & \textbf{Posterior mean} & \textbf{95\% CI} \\
\hline
\phantom{0}1 & 3--6 & 65.0 & 39.5 & $(27.9, 52.3)$ \\
\phantom{0}2 & 1--3 & 33.0 & 20.2 & $(12.1, 30.2)$ \\
\phantom{0}3 & 5--2 & 67.0 & 59.5 & $(42.7, 78.7)$ \\
\phantom{0}4 & 6--1 & 38.0 & 26.4 & $(16.3, 38.4)$ \\
\phantom{0}5 & 3--1 & 28.0 & 21.9 & $(12.4, 33.1)$ \\
\phantom{0}6 & 2--6 & 30.0 & 38.3 & $(25.2, 53.1)$ \\
\phantom{0}7 & 6--2 & 37.0 & 59.8 & $(41.8, 78.9)$ \\
\phantom{0}8 & 1--5 & \phantom{0}9.0 & \phantom{0}5.5 & $\phantom{0}(1.7, 11.4)$ \\
\phantom{0}9 & 1--6 & 30.0 & 20.5 & $(11.8, 31.3)$ \\
10 & 2--5 & 37.0 & 33.8 & $(22.2, 47.5)$ \\
11 & 2--3 & 37.0 & 36.7 & $(26.2, 48.9)$ \\
12 & 3--5 & 20.0 & 10.7 & $\phantom{0}(5.1, 18.2)$ \\
13 & 3--2 & 20.0 & 51.6 & $(37.8, 67.0)$ \\
14 & 1--2 & \phantom{0}2.0 & 15.7 & $\phantom{0}(6.3, 26.3)$ \\
15 & 5--6 & 20.0 & 28.0 & $(16.5, 41.5)$ \\
16 & 2--1 & \phantom{0}2.0 & \phantom{0}6.4 & $\phantom{0}(0.3, 15.4)$ \\
17 & 6--5 & 31.0 & 66.7 & $(49.0, 86.2)$ \\
18 & 5--3 & 10.0 & \phantom{0}4.7 & \phantom{0}$(1.5, 9.5)$\phantom{0} \\
19 & 6--3 & 15.0 & \phantom{0}8.0 & \phantom{0}$(3.4, 14.3)$ \\
20 & 5--1 & 69.0 & 38.9 & $(27.7, 52.1)$ \\
\hline
\end{tabular*}
\end{table}

We conducted Bayesian inference for the mean route flows $\bolds
{\theta}$
using a Poisson route flow model,
$x_j \sim\operatorname{\mathsf{Pois}}(\theta_j)$ independently for
$j=1,\ldots,r$. As in the
previous application, the two pilot
phases of our MCMC sampler ran for $10{,}000$ iterations
each. The algorithm then ran for a further $20{,}000$ iterations. The
total computing time was slightly less
than $10$ minutes using the same computing environment as described in
the previous application.

%f9
%f9 #&#

%t3 #&#
\begin{table}[b]
\caption{Available link count data for the network in
Figure \protect\ref{fig9}}
\label{tab3}
\begin{tabular*}{\tablewidth}{@{\extracolsep{\fill}}lccccccccc@{}}
\hline
Link & \phantom{000}1 & \phantom{000}2 & \phantom{000}5 & \phantom{000}6 & \phantom{000}7 & \phantom{00}11 & \phantom{00}13 & \phantom{00}14 & \phantom{00}16 \\
Count & 1279 & \phantom{0}740 & 1112 & \phantom{0}826 & 1221 & 1147 & 1066 & \phantom{0}764 & \phantom{0}835 \\[3pt]
Link & \phantom{00}18 & \phantom{00}21 & \phantom{00}22 & \phantom{00}25 & \phantom{00}27 & \phantom{00}29 & \phantom{00}31 & \phantom{00}32 & \phantom{00}34 \\
Count & \phantom{0}462 & \phantom{0}137 & \phantom{0}193 & \phantom{0}746 & \phantom{0}685 & \phantom{0}466 & \phantom{0}538 & \phantom{0}499 & \phantom{0}453 \\[3pt]
Link & \phantom{00}35 & \phantom{00}36 & \phantom{00}37 & \phantom{00}38 & \phantom{00}39 & \phantom{00}40 & \phantom{00}42 & \phantom{00}46 & \phantom{00}47 \\
Count & \phantom{0}610 & \phantom{0}503 & \phantom{0}667 & \phantom{0}483 & \phantom{0}545 & \phantom{0}500 & \phantom{00}57 & \phantom{0}194 & \phantom{0}111 \\
\hline
\end{tabular*}
\end{table}

\begin{figure}

\includegraphics{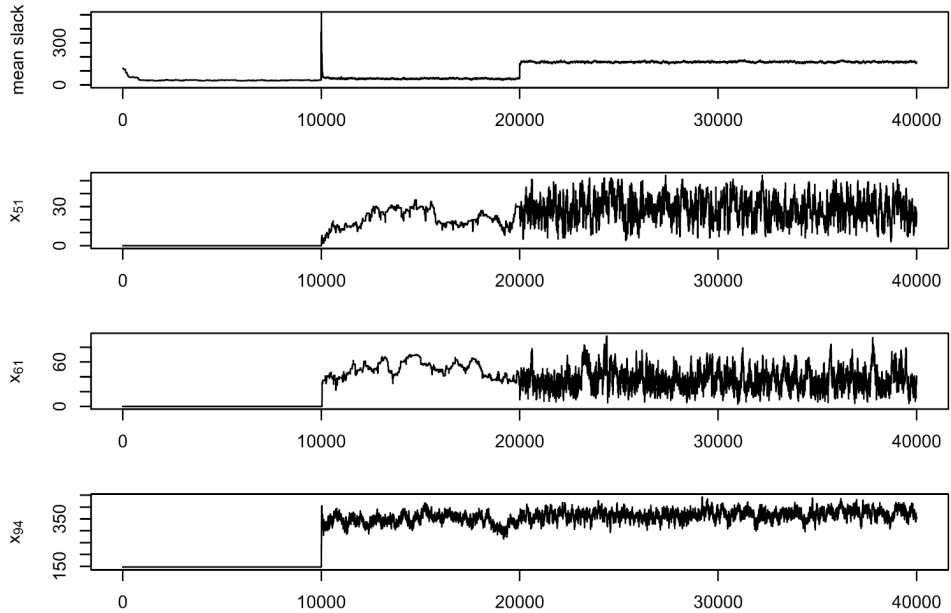}

\caption{Trace plots for sampled route flows for the
network in Figure~\protect\ref{fig9}. The top panel
displays the mean flow for routes corresponding to partition $A_1$, and
hence is a measure of slack. The remaining
three panels display sampled flows for three selected routes.}\label{fig10}
\end{figure}

In Figure~\ref{fig10} we display the trace plots for $x_j$ for three
selected routes. We also display the
mean flow on the routes corresponding to partition $A_1$ as a measure
of slack in the linear system
(\ref{eqfundamental}). The sampler is almost completely stuck during
the first pilot phase, with only 16 of
the 127 routes updated (in the sense of a change of value) at any stage
in the first 10{,}000 iterations. The situation
is much improved in the second pilot phase, with all route flows
updating. Nonetheless, the mixing of the sampler
during this phase is relatively poor. A second optimization of the
route ordering leads to better mixing
throughout the final 20{,}000 iterations of the algorithm.

As expected, the performance of the sampler at the various phases is
mirrored by the mean slack in the swap space.
To provide further insight into this effect, consider breaking down the
first 30{,}000 iterations into equally
sized blocks of 10{,}000 iterations. These correspond to the three
partitions of $A$ that are employed. Over the
first block the mean slack is $35.9$, over the second block the mean
slack is $46.4$, and over the third block the mean slack
is $163.2$. Using the final block as a benchmark, the sampling
efficiencies in the first two blocks (measured in terms
of mean effective sample size for the traffic flows on the three
selected routes) are 0\% and 13.5\%, respectively.

%f10 #&#

These results show that implementation of \citeauthor{TebaldiWest98}'s (\citeyear{TebaldiWest98}) algorithm without
modification of the route order
fails completely
to converge based on the initial partition of $A$. Moreover, even using
the route ordering during the second
pilot phase (which is partially optimized) gives a sampler with quite
poor mixing properties.

In order to check that these problems were not a consequence of a
pathological data set or initial route ordering, we
ran a small simulation study in which link counts were generated from a
Poisson model fitted using the posterior mean
route flows and where the columns of $A$ were ordered at random
(subject to the constraint that $A_1$ is invertible).
In 100 replications, the unmodified version of \citeauthor{TebaldiWest98}'s (\citeyear{TebaldiWest98}) algorithm failed to mix
on every occasion,
in the sense that there was at least one route flow that was not
updated. In contrast, use of our algorithm with
two updates of the partition of $A$ led to acceptable mixing in every
replication. This indicates that while there
are a very large number of possible partitions for $A$, it is necessary
to search carefully for ones that produce
a sampler with good properties.

\begin{figure}

\includegraphics{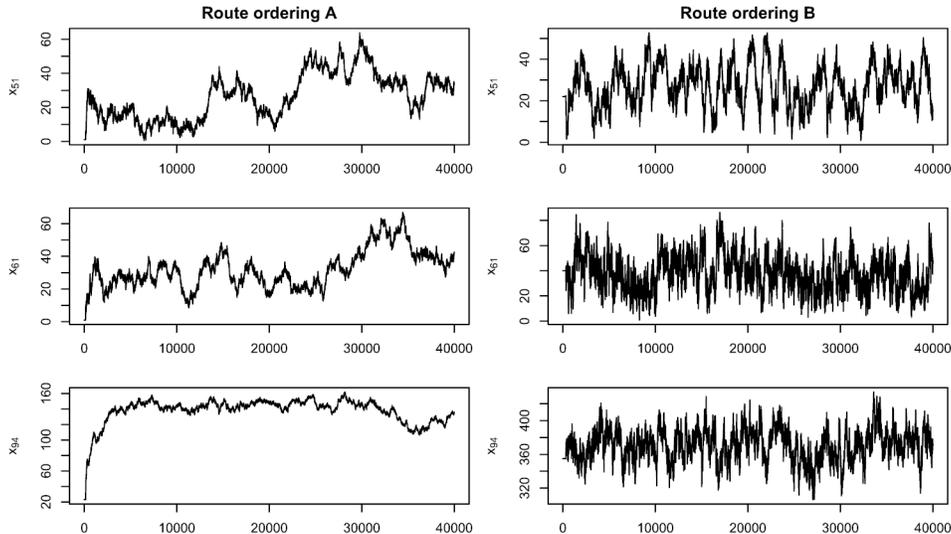}

\caption{Trace plots for sampled route flows for the
network in Figure~\protect\ref{fig9} using
the random search direction algorithm of \citet{AiroldiBlocker13}. The left-hand panels show results
obtained using route ordering A; those on the right-hand side show
results obtained using the route
ordering B.}\label{fig11}
\end{figure}

We also tried applying the random search direction sampler of
\citet{AiroldiBlocker13} to this application, based on
code harvested from the R library \textsf{networkTomography}
[\citet{BlockerKoullickAiroldi12}]. This methodology is
intended for continuous flows, and so we employed an approximation in
which the likelihoods in the acceptance
probability were computed using rounded versions of the sampled route
flows. We implemented this sampler using
two route orderings. The first (``ordering A'') matches that used for
the first pilot phase in our algorithm, while the second
(``ordering B'') is the final (optimized) route ordering found by our
algorithm. We thin the output, retaining only every
$(r-n)$th (i.e., 77th) iteration, to create a fair comparison with the
results from our algorithm (where one iteration
involves updating the flows on routes corresponding to all 77 columns
of $A_2$). Thinned trace places for flows on selected
routes (corresponding to those in Figure~\ref{fig10}) are given in
Figure~\ref{fig11}. The computing time was approximately
2.5 times slower than that for our algorithm.

%f11 #&#

It is evident from these results that \citeauthor{AiroldiBlocker13}'s
(\citeyear{AiroldiBlocker13}) algorithm works far better than the
\citet{TebaldiWest98} algorithm for the initial partition of $A$.
However, the trace plots also indicate that while the
properties of Airoldi and Blocker's sampler are improved through a
refined route ordering, the sampler mixes somewhat
less well than our algorithm with the optimized partition of $A$. This
result ties in with our examination of Example~\ref{exa3}
in Section~\ref{secsampling}. Componentwise sampling fails entirely,
or mixes extremely slowly, when the partition of
$A$ gives rise to a polytope with awkward geometry. However, when the
partition is updated to maximize the slack, then
``long moves'' are possible in the coordinate directions, and sampling
in those directions can be preferable to random
search directions. Further insight is provided by a comparison of the
slack for all the sampling algorithms considered.
See the supplementary material for details [\citet{H15aoas-supp}].

%s6 #&#
\section{Discussion}
\label{secdiscuss}

The availability of a dependable method for sampling route flows
conditional on an observed pattern
of link counts is pivotal to estimation of origin--destination traffic
volumes and associated statistical
network tomography problems. As we have shown, implementation of
\citeauthor{TebaldiWest98}'s (\citeyear{TebaldiWest98}) proposed
sampler with a fixed partition of the routing matrix is unreliable
because the polytope of feasible
route flows may be oriented at an awkward angle to the sampling
directions. Nonetheless, the difficulties
are resolved by a change of coordinate representation of the polytope
through a reordering of the
columns of $A$. Indeed, given that there is always a good route
ordering available (if $A$
is totally unimodular), componentwise sampling of the elements of
$\mathbf{x}
_2$ is adequate. This is fortunate,
since we speculate that it would be very difficult to develop an exact
sampler [as opposed to a continuous
approximation like that of \citet{AiroldiBlocker13}] to draw
candidate route flows from higher dimensional
spaces when the traffic is integer valued.

The unimodularity requirements on $A$ place a caveat on the preceding
remarks, although not a serious one.
In practice, we require only that a good route ordering is available
for which $A_1$ is unimodular (and
hence invertible as an integer-valued matrix). This is guaranteed if
$A$ is totally unimodular, but may
well occur even when this is not the case: most of the $A_1$
submatrices can be unimodular even if $A$
is not totally unimodular. It follows that total unimodularity is a
sufficient, but by no means necessary,
condition for the proposed sampling algorithm to work effectively.

\begin{figure}

\includegraphics{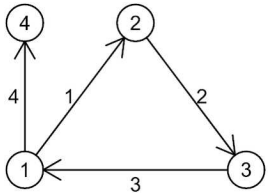}

\caption{A potentially problematic network. If traffic
counts are available for only links 1, 2 and 3,
and if travel is possible only between the ordered node pairs $(1,3)$,
$(2,1)$, $(3,2)$ and $(3,4)$, then the link-path
incidence matrix is not totally unimodular.}\label{fig12}
\end{figure}

The previous comments notwithstanding, it is still of interest to
explore further the issue of total
unimodularity. As \citet{AiroldiBlocker13} indicate, the routing
matrices that are encountered in practice
seem to be totally unimodular almost without exception. Nonetheless,
one does not have to work too hard to
find a network tomography problem for which this is not the case.
Consider the network displayed in Figure~\ref{fig12}. Suppose that traffic counts are observed on links 1, 2
and 3 only, and that travel is
possible for O--D pairs $(1,3)$, $(2,1)$, $(3,2)$ and $(3,4)$ via the
obvious acyclic routes.

%f12 #&#

Based on that route ordering, the link-path incidence matrix is given by
%
%e15 #&#
\begin{equation} \label{eqAnotTUM}
A = [A_1 | A_2 ] = \lleft[ %
\begin{array}
{c@{\quad}c@{\quad}c@{\hspace*{6pt}}|@{\hspace*{6pt}}c} 1 & 0 & 1 & 0
\\
1 & 1 & 0 & 0
\\
0 & 1 & 1 & 1 \end{array} %
 \rright].
\end{equation}
The submatrix $A$ is not unimodular [$\det(A_1) = 2$], and thus admits
a noninteger-valued
inverse matrix,
\[
A_1^{-1} = \lleft[ %
\begin{array}{c@{\quad}c@{\quad}c}
\tfrac{1}2 & \tfrac{1}2 & -\tfrac{1}2
\\[4pt]
-\tfrac{1}2 & \tfrac{1}2 & -\tfrac{1}2
\\[4pt]
\tfrac{1}2 & -\tfrac{1}2 & \tfrac{1}2 \end{array}
 \rright].
\]
An immediate consequence is that if we attempt to implement the conditional
route sampler using $A$ from (\ref{eqAnotTUM}), then the resultant
traffic flows need not be integer
valued. However, by switching any of the first three columns with the
fourth, we obtain new partitions
$A=[A_1 | A_2]$ where the elements of $A_1^{-1}$ lie in the set $\{-1,
0, 1\}$. Our route flow
sampler works successfully when only these partitions of $A$ are considered.

It is interesting to reflect on the characteristics of this problem
that led to the lack of total unimodularity
in (\ref{eqAnotTUM}). The routing scheme is unusual (and rather
artificial) because travel is only
possible over paths comprising exactly two links. Since traffic counts
are available for links 1, 2
and 3, the result is a submatrix $A_1$ in~(\ref{eqAnotTUM}) where all
the column sums equal two.
No routing submatrix with this property can be unimodular, a result
that can be generalized as
follows.

\begin{prop}\label{prop4}
If $A_1$ is unimodular, then its column sums are coprime.
\end{prop}

The proof is given in the \hyperref[app]{Appendix}.

When we reverse the final two columns of $A$, the column sums of the
resulting submatrix $A_1$
are coprime and $A_1$ is unimodular. Nonetheless, this coprime property
is insufficient to
ensure unimodularity in general. For example, suppose that we connect
the networks in Figures~\ref{fig1} and \ref{fig8} by inserting a link joining node 4 of the
former to node 3 of
the latter. If we leave the pattern of permissible O--D pairs and routes
unchanged, then
we may order the routes so that the submatrix $A_1$ for the composite
network is block
diagonal, with the $A_1$ matrices from the original matrices forming
the blocks. The combined
matrix $A_1$ will have coprime column totals, but will obviously not be
unimodular.

This example is somewhat extreme, however. While the composite network
is physically connected,
it operates as two independent (sub)networks since there are no routes
requiring travel from one
to the other. Whether unimodularity of $A_1$ can be guaranteed by
imposing some natural assumptions
on the routes of the network remains unclear.
\begin{appendix}
\section*{Appendix}\label{app}
\subsection*{Proof of Proposition~\protect\ref{prop1}}

Let $\mathcal{Y}$ denote the (unconditional) support of $\mathbf{y}$.
It suffices to
show that
$f_Y( \mathbf{y}| \bolds{\theta}) = f_Y( \mathbf{y}| \tilde{\bolds
{\theta}})$ for all $\mathbf{y}\in
\mathcal{Y}$ implies that
$f_X( \mathbf{x}| \bolds{\theta}) = f_X( \mathbf{x}| \tilde{\bolds
{\theta}})$ for all $\mathbf{x}\in
\mathcal{X}$. By the independence assumption on the
route flows, it is sufficient to show that $f_{X_j}(h | \bolds{\theta
}) =
f_{X_j}(h | \tilde{\bolds{\theta}})$ for all $h \in\mathbb{Z}_{\ge0}$
and $j \in\mathcal{R}$.

Assume henceforth that $f_Y( \mathbf{y}| \bolds{\theta}) = f_Y(
\mathbf{y}| \tilde{\bolds{\theta}
})$ for all $\mathbf{y}\in\mathcal{Y}$.
Then using the fact that all columns of $A$ contain at least one
nonzero element, we obtain
%
%e16 #&#
\begin{equation}\label{eqapp1}
f_X(\mathbf{0} | \bolds{\theta}) = f_Y(\mathbf{0} |
\bolds{\theta }) = f_Y(\mathbf{0} | \tilde{\bolds{\theta}}) =
f_X(\mathbf{0} | \tilde{\bolds{\theta}}),
\end{equation}
where $\mathbf{0}$ denotes an appropriately dimensioned vector of zeroes.

The proof now proceeds by induction on route length. To this end, we
define $s_j$ to be the number of constituent monitored
links for route $j \in\mathcal{R}$, that is, $s_j = \Vert \mathbf{a}_j
\Vert_1 = \sum_{i=1}^c a_{ij}$, where $\mathbf{a}_j$ denotes the $j$th column
of $A$. Let $s_{(1)} < s_{(2)} < \cdots < s_{(\rho)}$ denote the unique
values of $s_j$ in increasing order. Now partition the
routes into (disjoint) sets $J_1, J_2, \ldots, J_\rho$, where $J_i$
contains all routes comprising $s_{(i)}$ links.

Consider now $\mathbf{x}= h \mathbf{e}_j$ for $j \in J_1$, where
$\mathbf{e}_j$ is
the $j$th coordinate vector (i.e., a~vector of zeros
except for a one in the $j$th position) and $h \in\mathbb{Z}_{\ge1}$.
This is a route flow pattern with $h$ vehicles on
route $j$ and none on any other route. Then
\begin{eqnarray*}
\frac{f_{X_j}(h | \bolds{\theta})}{f_{X_j}(0 | \bolds{\theta})} f_X(\mathbf{0} | \bolds{\theta}) &=&
f_X(h \mathbf{e}_j | \bolds{\theta})
\\
&=& f_Y(h \mathbf{a}_j | \bolds{\theta})
\\
&=& f_Y(h \mathbf{a}_j | \tilde{\bolds{\theta}})
\\
&=& f_X(h \mathbf{e}_j | \tilde{\bolds{\theta}})
\\
&=& \frac{f_{X_i}(h | \tilde{\bolds{\theta}})}{f_{X_i}(0 | \tilde
{\bolds{\theta}})} f_X(\mathbf{0} | \tilde{\bolds{\theta}}),
\end{eqnarray*}
where the penultimate equality holds because $\mathbf{x}= h \mathbf
{e}_j$ is the
unique solution of $A \mathbf{x}= h \mathbf{a}_j$.
To see this, note that there can be no solution involving shorter
routes (there are none) or longer routes (since
this would be incompatible with a link flow pattern involving just
$s_{(1)}$ links), and there can be no solution
involving a route $j' \ne j \in J_1$ because the columns of $A$ are
distinct. Applying equation (\ref{eqapp1}), it
follows that
%
%e17 #&#
\begin{equation} \label{eqapp2}
f_{X_j}(h | \bolds{\theta}) f_{X_j}(0 | \tilde{\bolds{\theta}}
) = f_{X_j}(h | \tilde {\bolds{\theta}}) f_{X_j}(0 | \bolds{
\theta})
\end{equation}
for $h =1,2,3 \ldots.$ Equation (\ref{eqapp2}) also holds trivially
when $h=0$. We may therefore sum
(\ref{eqapp2}) over $h \in\mathbb{Z}_{\ge0}$ to give
\[
f_{X_j}(0 | \tilde{\bolds{\theta}} ) = f_{X_j}(0 | \bolds{
\theta}).
\]
It follows immediately from (\ref{eqapp2}) that for $j \in J_1$,
$f_{X_j}(h | \tilde{\bolds{\theta}} ) = f_{X_j}(h | \bolds{\theta})$
for all $h \in\mathbb{Z}_{\ge0}$.

For the purposes of induction, assume now that $f_{X_j}(h | \tilde
{\bolds{\theta}} ) = f_{X_j}(h | \bolds{\theta})$ for all $h \in
\mathbb{Z}_{\ge0}$
for all $j \in J_1 \cup J_2 \cup\cdots\cup J_k$. Consider $\mathbf
{x}= h
\mathbf{e}_j$ for $j \in J_{k+1}$.

Starting with $h=1$, we have
\[
f_Y(\mathbf{a}_j | \bolds{\theta}) = f_X(
\mathbf{e}_j | \bolds {\theta}) + \mathop{\sum
_{\mathbf{x} \ne\mathbf{e}_j}}_{A\mathbf{x}= \mathbf{a}_j} f_X(\mathbf{x}| \bolds{
\theta}).
\]
Now, if $\mathbf{x}\ne\mathbf{e}_j$ satisfies $A\mathbf{x}= \mathbf
{a}_j$, then $x_j = 0$
for all $j \in J_i$ for $i \ge k$ [using exactly
the same kind of argument as preceded equation (\ref{eqapp2})]. It
follows that, for any such~$\mathbf{x}$,
%
%e18 #&#
\begin{equation}\label{eqapp3}
f_X(\mathbf{x}| \bolds{\theta}) = f_X(\mathbf{0} |
\bolds{\theta }) \prod_{i \in J^\dagger} \frac{f_{X_i}(1 | \bolds{\theta})}{f_{X_i}(0 | \bolds{\theta})}
\end{equation}
for some indexing set $J^\dagger\subseteq J_1 \cup\cdots\cup J_k$.
There\vspace*{1pt} is no need to specify this set precisely: it
is sufficient to know that for $j \in J^\dagger$, $f_{X_j}(h | \bolds
{\theta})
= f_{X_j}(h | \tilde{\bolds{\theta}})$ for all $h$ by
our inductive hypothesis.

Hence, from equation (\ref{eqapp3}),
\begin{eqnarray*}
f_Y(\mathbf{a}_j | \bolds{\theta}) &=&
f_X(\mathbf{e}_j | \bolds {\theta}) + \mathop{\sum
_{\mathbf{x}\ne\mathbf{e}_j}}_{A\mathbf{x}= \mathbf{a}_j } f_X(\mathbf{x}|
\bolds{\theta})
\\
&=& f_X(\mathbf{e}_j | \bolds{\theta}) + \mathop{\sum
_{\mathbf{x}\ne\mathbf{e}_j}}_{A\mathbf{x}
= \mathbf{a}_j} f_X(\mathbf{x}|
\tilde{\bolds{\theta}})
\\
&=& f_X(\mathbf{e}_j | \bolds{\theta}) +
f_Y(\mathbf{a}_j | \tilde{\bolds{\theta}}) -
f_X(\mathbf{e}_j | \tilde{\bolds{\theta}}).
\end{eqnarray*}
Since $f_Y(\mathbf{a}_j | \bolds{\theta}) = f_Y(\mathbf{a}_j |
\tilde{\bolds{\theta}})$, it
follows that $f_X(\mathbf{e}_j | \bolds{\theta}) = f_X(\mathbf{e}_j
| \tilde{\bolds{\theta}})$
and, therefore,
\[
\frac{f_{X_j}(1 | \bolds{\theta})}{f_{X_j}(0 | \bolds{\theta})} f_X(\mathbf{0} | \bolds{\theta}) =
\frac{f_{X_j}(1 | \tilde{\bolds{\theta}})}{f_{X_j}(0 | \tilde
{\bolds{\theta}})} f_X(\mathbf{0} | \tilde{\bolds{\theta}})
\]
and so
%
%e19 #&#
\begin{equation}\label{eqapp4}
\frac{f_{X_j}(1 | \bolds{\theta})}{f_{X_j}(0 | \bolds{\theta})} = \frac{f_{X_j}(1 |
\tilde{\bolds{\theta}})}{f_{X_j}(0 | \tilde{\bolds{\theta}})},
\end{equation}
courtesy of equation (\ref{eqapp1}).

We continue by applying another ``inner'' mathematical induction, to
demonstrate that equation (\ref{eqapp4}) applies for
flows $h > 1$. For the purposes of induction, assume that
\[
\frac{f_{X_j}(h^* | \bolds{\theta})}{f_{X_j}(0 | \bolds{\theta})} = \frac{f_{X_j}(h^*
| \tilde{\bolds{\theta}})}{f_{X_j}(0 | \tilde{\bolds{\theta}})}
\]
for all $h^* = 1,2,\ldots,(h-1)$.

Now,
%
%e20 #&#
\begin{equation}\label{eqapp5}
f_Y(h \mathbf{a}_j | \bolds{\theta}) =
f_X(h \mathbf{e}_j | \bolds {\theta}) + \mathop{\sum
_{\mathbf{x}\ne h \mathbf{e}_j}}_{A\mathbf{x}= h \mathbf{a}_j} f_X(\mathbf{x}|
\bolds{\theta}).
\end{equation}
For every $\mathbf{x}\ne h \mathbf{e}_j$ such that $A\mathbf{x}= h
\mathbf{a}_j$,
%
%e21 #&#
\begin{equation}\label{eqapp6}
f_X(\mathbf{x}| \bolds{\theta}) = f_X(\mathbf{0} |
\bolds{\theta }) \prod_{h_0^* = 0}^{h-1}
\frac{f_{X_j}(h_0^* | \bolds{\theta})}{f_{X_j}(0 | \bolds{\theta})} \prod_{i \in J^\dagger} \frac{f_{X_i}(h_i^* | \bolds{\theta
})}{f_{X_i}(0 |
\bolds{\theta})},
\end{equation}
where $\{ h_j^* \}$ is a set of positive integers no greater than $h -
h_0^*$, and $J^\dagger\subseteq J_1 \cup\cdots\cup J_k$
(again requiring no explicit specification). Intuitively, this equation
relies on the fact that the link flow pattern $h \mathbf{a}_j$
can be generated by placing $h_0^*$ vehicles on route~$j$ and then
splicing together flows on compatible shorter routes to account
for the remaining $h - h_0^*$ vehicles.

The inductive\vspace*{1pt} hypotheses imply that equality is maintained in equation
(\ref{eqapp6}) if $\bolds{\theta}$ is replaced\vspace*{1pt} by
$\tilde{\bolds{\theta}}$ everywhere on the right-hand side. It
follows that
$f_X(\mathbf{x}| \bolds{\theta}) = f_X(\mathbf{x}| \tilde{\bolds
{\theta}})$,
when from (\ref{eqapp5}) we obtain
\begin{eqnarray*}
f_Y(h \mathbf{a}_j | \bolds{\theta}) &=&
f_X(h \mathbf{e}_j | \bolds{\theta}) + \mathop{\sum
_{\mathbf{x}\ne h \mathbf{e}_j}}_{A\mathbf{x}= h \mathbf
{a}_j} f_X(\mathbf{x}|
\bolds{\theta} )
\\
&=& f_X(h \mathbf{e}_j | \bolds{\theta}) + \mathop{
\sum_{\mathbf{x}\ne h \mathbf{e}_j }}_{A\mathbf{x}= h \mathbf{a}_j} f_X(
\mathbf{x}| \tilde{\bolds {\theta}})
\\
&=& f_X(h\mathbf{e}_j | \bolds{\theta}) +
f_Y(h\mathbf{a}_j | \tilde{\bolds{\theta}}) -
f_X(h\mathbf{e}_j | \tilde{\bolds{\theta}})
\\
&=& f_X(h\mathbf{e}_j | \bolds{\theta}) +
f_Y(h\mathbf{a}_j | \bolds{\theta}) -
f_X(h\mathbf{e} _j | \tilde{\bolds{\theta}}).
\end{eqnarray*}
It follows that $f_X(h\mathbf{e}_j | \bolds{\theta}) = f_X(h\mathbf
{e}_j | \tilde
{\bolds{\theta}})$ and so
\[
\frac{f_{X_j}(h | \bolds{\theta})}{f_{X_j}(0 | \bolds{\theta})} = \frac{f_{X_j}(h |
\tilde{\bolds{\theta}})}{f_{X_j}(0 | \tilde{\bolds{\theta}})},
\]
completing the inner inductive step.

We have proved that $f_{X_j}(h | \bolds{\theta}) f_{X_j}(0 | \tilde
{\bolds{\theta}})
= f_{X_j}(h | \tilde{\bolds{\theta}}) f_{X_j}(0 | \bolds{\theta})$
for all $h \in
\mathbb{Z}_{\ge0}$. Summing over $h$ gives $f_{X_j}(0 | \tilde
{\bolds{\theta}
}) = f_{X_j}(0 | \bolds{\theta})$ when it follows that
$f_{X_j}(h | \bolds{\theta}) = f_{X_j}(h | \tilde{\bolds{\theta
}})$ for all $j \in
J_{k+1}$. This completes the outer mathematical induction.

We conclude that $f_{X_j}(h | \bolds{\theta}) = f_{X_j}(h | \tilde
{\bolds{\theta}})$
for all $j \in\mathcal{R}$ and $h \in
\mathbb{Z}_{\ge0}$, completing the proof of Proposition~\ref{prop1}.

\subsection*{Proof of Proposition~\protect\ref{prop3}}

By Lemma~2.2 of \citet{AiroldiHass11}, the matrix $U$ is totally
unimodular, and therefore all its entries
lie in $\{-1,0,1\}$. Hence, if $\mathbf{u}_{1,j}$ is the vector formed from
the first $n$ elements of $\mathbf{u}_j$, we
have $\mathbf{x}_1 + \mathbf{u}_{1,j} \ge0$ (interpreted
componentwise), and hence
$\mathbf{x}+ \mathbf{u}_j \in\mathcal{X}_{|\mathbf{y}}$ as required.

\subsection*{Proof of Proposition~\protect\ref{prop4}}

Let $A_1$ be unimodular, and let $\mathbf{a}_*$ be the vector of column
sums of this matrix.
Suppose that the elements of $\mathbf{a}_*$ are not coprime and so
have a
greatest common
divisor of $d > 1$. Then the elements of the vector $d^{-1} \mathbf
{a}_*^\mathsf{T}
A_1^{-1}$ are
integers because $A_1^{-1}$ is an integer-valued matrix. However,
\[
d^{-1} \mathbf{a}_*^\mathsf{T}A_1^{-1} =
d^{-1} {\boldsymbol {1}}^\mathsf{T}A_1
A_1^{-1} = d^{-1} {\boldsymbol{1}}^\mathsf{T},
\]
providing a contradiction, and hence proving the result.
\end{appendix}

\section*{Acknowledgments}

The author acknowledges funding for the initial stages of this work
from the Royal Society of New Zealand
Marsden Grant scheme and a short but helpful conversation on unimodular
matrices with Dr. Chris Tuffley
(Massey University, NZ) and Professor Terence Tao (UCLA). The author
also thanks two anonymous referees
and the Editor for their helpful comments.

\begin{supplement}[id=suppA]
\stitle{Supplement to ``Network tomography for integer-valued traffic''}
\slink[doi]{10.1214/15-AOAS805SUPP} %[doi,text={...}] - jei reikia
%suskaldyti doi
\sdatatype{.pdf}
\sfilename{aoas805\_supp.pdf}
\sdescription{The supplementary materials, stored as a zip archive,
include data and additional numerical results for the applications in
Section~\ref{secapplication}.
The data comprise link-path incidence matrices, observed traffic counts
and prior pseudo counts for Bayesian analyses. The
additional results include effective sample sizes for MCMC output,
computing times and summaries of the slack for the
route flow samplers considered.}
\end{supplement}

% imsref loaded by daiva.urboniene, 2015-01-27 13:06:40

%


\printaddresses

\begin{thebibliography}{36}
%b1 ###
%b1 #&#
\bibitem[\protect\citeauthoryear{Airoldi and Blocker}{2013}]{AiroldiBlocker13}
\begin{barticle}[mr]
\bauthor{\bsnm{Airoldi},~\bfnm{Edoardo~M.}\binits{E.~M.}} \AND
\bauthor{\bsnm{Blocker},~\bfnm{Alexander~W.}\binits{A.~W.}}
(\byear{2013}).
\btitle{Estimating latent processes on a network from indirect measurements}.
\bjournal{J. Amer. Statist. Assoc.}
\bvolume{108}
\bpages{149--164}.
\bid{doi={10.1080/01621459.2012.756328}, issn={0162-1459}, mr={3174609}}
\end{barticle}
%

\bptok{imsref}%
% NOT OUTPUTTED:
%   number = 501
%   doi = http://dx.doi.org/10.1080/01621459.2012.756328
%   fjournal = Journal of the American Statistical Association
\endbibitem

%b2 ###
%b2 #&#
\bibitem[\protect\citeauthoryear{Airoldi and Haas}{2011}]{AiroldiHass11}
\begin{binproceedings}[author]
\bauthor{\bsnm{Airoldi},~\bfnm{E.~M.}\binits{E.~M.}} \AND
\bauthor{\bsnm{Haas},~\bfnm{B.}\binits{B.}}
(\byear{2011}).
\btitle{Polytope samplers for inference in ill-posed inverse problems}.
In \bbooktitle{International Conference on Artificial Intelligence and Statistics 15}.
\blocation{Ft. Lauderdale, FL}.
\end{binproceedings}
%

\bptok{imsref}%
\endbibitem

%b3 ###
%b3 #&#
\bibitem[\protect\citeauthoryear{Bell}{1991}]{Bell91}
\begin{barticle}[mr]
\bauthor{\bsnm{Bell},~\bfnm{Michael~G.~H.}\binits{M.~G.~H.}}
(\byear{1991}).
\btitle{The estimation of origin-destination matrices by constrained generalised least squares}.
\bjournal{Transp. Res. Part B}
\bvolume{25}
\bpages{13--22}.
\bid{doi={10.1016/0191-2615(91)90010-G}, issn={0191-2615}, mr={1093618}}
\end{barticle}
%

\bptok{imsref}%
% NOT OUTPUTTED:
%   number = 1
%   doi = http://dx.doi.org/10.1016/0191-2615(91)90010-G
%   coden = TRBMDY
%   fjournal = Transportation Research. Part B. Methodological. An International Journal
\endbibitem

%b4 ###
%b4 #&#
\bibitem[\protect\citeauthoryear{Ben-Akiva and Lerman}{1985}]{BenAkivaLerman85}
\begin{bbook}[author]
\bauthor{\bsnm{Ben-Akiva},~\bfnm{M.}\binits{M.}} \AND
\bauthor{\bsnm{Lerman},~\bfnm{S.~R.}\binits{S.~R.}}
(\byear{1985}).
\btitle{Discrete Choice Analysis}.
\bpublisher{MIT Press},
\blocation{Cambridge, MA.}
\end{bbook}
%

\bptok{imsref}%
\endbibitem

%b5 ###
%b5 #&#
\bibitem[\protect\citeauthoryear{Blocker, Koullick and Airoldi}{2012}]{BlockerKoullickAiroldi12}
\begin{bmisc}[author]
\bauthor{\bsnm{Blocker},~\bfnm{Alexander~W.}\binits{A.~W.}},
\bauthor{\bsnm{Koullick},~\bfnm{Paul}\binits{P.}} \AND
\bauthor{\bsnm{Airoldi},~\bfnm{Edoardo}\binits{E.}}
(\byear{2012}).
\bhowpublished{networkTomography: Tools for network tomography.
R package version 0.2.}
\end{bmisc}
%

\bptok{imsref}%
\endbibitem

%b6 ###
%b6 #&#
\bibitem[\protect\citeauthoryear{Caffo, Jank and Jones}{2005}]{CaffoJankJones05}
\begin{barticle}[mr]
\bauthor{\bsnm{Caffo},~\bfnm{Brian~S.}\binits{B.~S.}},
\bauthor{\bsnm{Jank},~\bfnm{Wolfgang}\binits{W.}} \AND
\bauthor{\bsnm{Jones},~\bfnm{Galin~L.}\binits{G.~L.}}
(\byear{2005}).
\btitle{Ascent-based {M}onte {C}arlo expectation-maximization}.
\bjournal{J. R. Stat. Soc. Ser. B Stat. Methodol.}
\bvolume{67}
\bpages{235--251}.
\bid{doi={10.1111/j.1467-9868.2005.00499.x}, issn={1369-7412}, mr={2137323}}
\end{barticle}
%

\bptok{imsref}%
% NOT OUTPUTTED:
%   number = 2
%   doi = http://dx.doi.org/10.1111/j.1467-9868.2005.00499.x
%   fjournal = Journal of the Royal Statistical Society. Series B. Statistical Methodology
\endbibitem

%b7 ###
%b7 #&#
\bibitem[\protect\citeauthoryear{Cao et~al.}{2000}]{CaoDavisVanderWielYu00}
\begin{barticle}[mr]
\bauthor{\bsnm{Cao},~\bfnm{Jin}\binits{J.}},
\bauthor{\bsnm{Davis},~\bfnm{Drew}\binits{D.}},
\bauthor{\bsnm{Vander Wiel},~\bfnm{Scott}\binits{S.}} \AND
\bauthor{\bsnm{Yu},~\bfnm{Bin}\binits{B.}}
(\byear{2000}).
\btitle{Time-varying network tomography: Router link data}.
\bjournal{J. Amer. Statist. Assoc.}
\bvolume{95}
\bpages{1063--1075}.
\bid{doi={10.2307/2669743}, issn={0162-1459}, mr={1821715}}
\end{barticle}
%

\bptok{imsref}%
% NOT OUTPUTTED:
%   number = 452
%   doi = http://dx.doi.org/10.2307/2669743
%   coden = JSTNAL
%   fjournal = Journal of the American Statistical Association
\endbibitem

%b8 ###
%b8 #&#
\bibitem[\protect\citeauthoryear{Cascetta}{1984}]{Cascetta84}
\begin{barticle}[author]
\bauthor{\bsnm{Cascetta},~\bfnm{E.}\binits{E.}}
(\byear{1984}).
\btitle{Estimation of trip matrices from traffic counts and survey data: A generalized least squares estimator}.
\bjournal{Transportation Research Part B}
\bvolume{18}
\bpages{289--299}.
\end{barticle}
%

\bptok{imsref}%
\endbibitem

%b9 ###
%b9 #&#
\bibitem[\protect\citeauthoryear{Cascetta}{1989}]{Cascetta89}
\begin{barticle}[author]
\bauthor{\bsnm{Cascetta},~\bfnm{E.}\binits{E.}}
(\byear{1989}).
\btitle{A stochastic process approach to the analysis of temporal dynamics in transportation networks}.
\bjournal{Transportation Research Part B}
\bvolume{23}
\bpages{1--17}.
\end{barticle}
%

\bptok{imsref}%
\endbibitem

%b10 ###
%b10 #&#
\bibitem[\protect\citeauthoryear{Cascetta et~al.}{1996}]{Cascettaetal96}
\begin{binproceedings}[author]
\bauthor{\bsnm{Cascetta},~\bfnm{E.}\binits{E.}},
\bauthor{\bsnm{Nuzzolo},~\bfnm{A.}\binits{A.}},
\bauthor{\bsnm{Russo},~\bfnm{F.}\binits{F.}} \AND
\bauthor{\bsnm{Vitetta},~\bfnm{A.}\binits{A.}}
(\byear{1996}).
\btitle{A modified logit route choice model overcoming path overlapping problems: Specification and some calibration results for interurban networks}.
In \bbooktitle{Proceedings of the 13th International Symposium on Transportation and Traffic Theory}
\bpages{697--711}.
\bpublisher{Elsevier Science},
\blocation{Lyon, France}.
\end{binproceedings}
%

\bptok{imsref}%
\endbibitem

%b11 ###
%b11 #&#
\bibitem[\protect\citeauthoryear{Castro et~al.}{2004}]{CastroCoatesLiangNowakYu04}
\begin{barticle}[mr]
\bauthor{\bsnm{Castro},~\bfnm{Rui}\binits{R.}},
\bauthor{\bsnm{Coates},~\bfnm{Mark}\binits{M.}},
\bauthor{\bsnm{Liang},~\bfnm{Gang}\binits{G.}},
\bauthor{\bsnm{Nowak},~\bfnm{Robert}\binits{R.}} \AND
\bauthor{\bsnm{Yu},~\bfnm{Bin}\binits{B.}}
(\byear{2004}).
\btitle{Network tomography: Recent developments}.
\bjournal{Statist. Sci.}
\bvolume{19}
\bpages{499--517}.
\bid{doi={10.1214/088342304000000422}, issn={0883-4237}, mr={2185628}}
\end{barticle}
%

\bptok{imsref}%
% NOT OUTPUTTED:
%   number = 3
%   doi = http://dx.doi.org/10.1214/088342304000000422
%   fjournal = Statistical Science. A Review Journal of the Institute of Mathematical Statistics
\endbibitem

%b12 ###
%b12 #&#
\bibitem[\protect\citeauthoryear{Daganzo and Sheffi}{1977}]{DaganzoSheffi77}
\begin{barticle}[author]
\bauthor{\bsnm{Daganzo},~\bfnm{C.~F.}\binits{C.~F.}} \AND
\bauthor{\bsnm{Sheffi},~\bfnm{Y.}\binits{Y.}}
(\byear{1977}).
\btitle{On stochastic models of traffic assignment}.
\bjournal{Transp. Sci.}
\bvolume{11}
\bpages{253--274}.
\end{barticle}
%

\bptok{imsref}%
\endbibitem

%b13 ###
%b13 #&#
\bibitem[\protect\citeauthoryear{Denby et~al.}{2007}]{Denbyetal07}
\begin{barticle}[mr]
\bauthor{\bsnm{Denby},~\bfnm{Lorraine}\binits{L.}},
\bauthor{\bsnm{Landwehr},~\bfnm{James~M.}\binits{J.~M.}},
\bauthor{\bsnm{Mallows},~\bfnm{Colin~L.}\binits{C.~L.}},
\bauthor{\bsnm{Meloche},~\bfnm{Jean}\binits{J.}},
\bauthor{\bsnm{Tuck},~\bfnm{John}\binits{J.}},
\bauthor{\bsnm{Xi},~\bfnm{Bowei}\binits{B.}},
\bauthor{\bsnm{Michailidis},~\bfnm{George}\binits{G.}} \AND
\bauthor{\bsnm{Nair},~\bfnm{Vijayan~N.}\binits{V.~N.}}
(\byear{2007}).
\btitle{Statistical aspects of the analysis of data networks}.
\bjournal{Technometrics}
\bvolume{49}
\bpages{318--334}.
\bid{doi={10.1198/004017007000000290}, issn={0040-1706}, mr={2408636}}
\end{barticle}
%

\bptok{imsref}%
% NOT OUTPUTTED:
%   number = 3
%   doi = http://dx.doi.org/10.1198/004017007000000290
%   coden = TCMTA2
%   fjournal = Technometrics. A Journal of Statistics for the Physical, Chemical and Engineering Sciences
\endbibitem

%b14 ###
%b14 #&#
\bibitem[\protect\citeauthoryear{Hazelton}{2001}]{H01jrssc}
\begin{barticle}[mr]
\bauthor{\bsnm{Hazelton},~\bfnm{Martin~L.}\binits{M.~L.}}
(\byear{2001}).
\btitle{Estimation of origin-destination trip rates in {L}eicester}.
\bjournal{J. Roy. Statist. Soc. Ser. C}
\bvolume{50}
\bpages{423--433}.
\bid{doi={10.1111/1467-9876.00245}, issn={0035-9254}, mr={1871797}}
\end{barticle}
%

\bptok{imsref}%
% NOT OUTPUTTED:
%   number = 4
%   doi = http://dx.doi.org/10.1111/1467-9876.00245
%   fjournal = Journal of the Royal Statistical Society. Series C. Applied Statistics
\endbibitem

%b15 ###
%b15 #&#
\bibitem[\protect\citeauthoryear{Hazelton}{2010}]{H10tech}
\begin{barticle}[mr]
\bauthor{\bsnm{Hazelton},~\bfnm{Martin~L.}\binits{M.~L.}}
(\byear{2010}).
\btitle{Statistical inference for transit system origin-destination matrices}.
\bjournal{Technometrics}
\bvolume{52}
\bpages{221--230}.
\bid{doi={10.1198/TECH.2010.09021}, issn={0040-1706}, mr={2757217}}
\end{barticle}
%

\bptok{imsref}%
% NOT OUTPUTTED:
%   number = 2
%   doi = http://dx.doi.org/10.1198/TECH.2010.09021
%   coden = TCMTA2
%   fjournal = Technometrics. A Journal of Statistics for the Physical, Chemical and Engineering Sciences
\endbibitem

%b16 ###
%b16 #&#
\bibitem[\protect\citeauthoryear{Hazelton}{2015}]{H15aoas-supp}
%
\begin{bmisc}[author]
\bauthor{\bsnm{Hazelton},~\bfnm{M.~L.}\binits{M.~L.}}
(\byear{2015}).
\bhowpublished{Supplement to ``Network tomography for integer-valued traffic.''
DOI:\doiurl{10.1214/15-AOAS805SUPP}}.
\bptok{imsref}%
\end{bmisc}
%
\endbibitem

%b17 ###
%b17 #&#
\bibitem[\protect\citeauthoryear{Heaton et~al.}{2012}]{Heatonetal12}
\begin{barticle}[author]
\bauthor{\bsnm{Heaton},~\bfnm{Luke}\binits{L.}},
\bauthor{\bsnm{Obara},~\bfnm{Boguslaw}\binits{B.}},
\bauthor{\bsnm{Grau},~\bfnm{Vincente}\binits{V.}},
\bauthor{\bsnm{Jones},~\bfnm{Nick}\binits{N.}},
\bauthor{\bsnm{Nakagaki},~\bfnm{Toshiyuki}\binits{T.}},
\bauthor{\bsnm{Boddy},~\bfnm{Lynne}\binits{L.}} \AND
\bauthor{\bsnm{Fricker},~\bfnm{Mark~D.}\binits{M.~D.}}
(\byear{2012}).
\btitle{Analysis of fungal networks}.
\bjournal{Fungal Biology Reviews}
\bvolume{26}
\bpages{12--29}.
\end{barticle}
%

\bptok{imsref}%
\endbibitem

%b18 ###
%b18 #&#
\bibitem[\protect\citeauthoryear{Hoffman and Kruskal}{1956}]{HoffmanKruskal56}
\begin{bincollection}[mr]
\bauthor{\bsnm{Hoffman},~\bfnm{A.~J.}\binits{A.~J.}} \AND
\bauthor{\bsnm{Kruskal},~\bfnm{J.~B.}\binits{J.~B.}}
(\byear{1956}).
\btitle{Integral boundary points of convex polyhedra}.
In \bbooktitle{Linear Inequalities and Related Systems}
(\beditor{\bfnm{H.~W.}\binits{H.~W.}~\bsnm{Kuhn}} \AND
\beditor{\bfnm{A.~W}\binits{A.~W.}~\bsnm{Tucker}}, eds.)
\bpages{223--246}.
\bpublisher{Princeton Univ. Press},
\blocation{Princeton, NJ}.
\bid{mr={0085148}}
\end{bincollection}
%

\bptok{imsref}%
\endbibitem

%b19 ###
%b19 #&#
\bibitem[\protect\citeauthoryear{Kolaczyk}{2009}]{Kolaczyk09}
\begin{bbook}[mr]
\bauthor{\bsnm{Kolaczyk},~\bfnm{Eric~D.}\binits{E.~D.}}
(\byear{2009}).
\btitle{Statistical Analysis of Network Data: Methods and Models}.
\bpublisher{Springer},
\blocation{New York}.
\bid{doi={10.1007/978-0-387-88146-1}, mr={2724362}}
\end{bbook}
%

\bptok{imsref}%
% NOT OUTPUTTED:
%   doi = http://dx.doi.org/10.1007/978-0-387-88146-1
%   isbn = 978-0-387-88145-4
%   fpage = xii+386
\endbibitem

%b20 ###
%b20 #&#
\bibitem[\protect\citeauthoryear{Koppelman and Wen}{2000}]{KoppelmanWen00}
\begin{barticle}[author]
\bauthor{\bsnm{Koppelman},~\bfnm{F.~S.}\binits{F.~S.}} \AND
\bauthor{\bsnm{Wen},~\bfnm{C.~H.}\binits{C.~H.}}
(\byear{2000}).
\btitle{The paired combinatorial logit model: Properties, estimation and application}.
\bjournal{Transp. Res., Part B: Methodol.}
\bvolume{34}
\bpages{75--89}.
\end{barticle}
%

\bptok{imsref}%
\endbibitem

%b21 ###
%b21 #&#
\bibitem[\protect\citeauthoryear{Lawrence, Michailidis and Nair}{2006}]{LawrenceMichailidisNair06}
\begin{barticle}[mr]
\bauthor{\bsnm{Lawrence},~\bfnm{Earl}\binits{E.}},
\bauthor{\bsnm{Michailidis},~\bfnm{George}\binits{G.}} \AND
\bauthor{\bsnm{Nair},~\bfnm{Vijayan~N.}\binits{V.~N.}}
(\byear{2006}).
\btitle{Network delay tomography using flexicast experiments}.
\bjournal{J. R. Stat. Soc. Ser. B Stat. Methodol.}
\bvolume{68}
\bpages{785--813}.
\bid{doi={10.1111/j.1467-9868.2006.00567.x}, issn={1369-7412}, mr={2301295}}
\end{barticle}
%

\bptok{imsref}%
% NOT OUTPUTTED:
%   number = 5
%   doi = http://dx.doi.org/10.1111/j.1467-9868.2006.00567.x
%   fjournal = Journal of the Royal Statistical Society. Series B. Statistical Methodology
\endbibitem

%b22 ###
%b22 #&#
\bibitem[\protect\citeauthoryear{Li}{2005}]{Li05}
\begin{barticle}[mr]
\bauthor{\bsnm{Li},~\bfnm{Baibing}\binits{B.}}
(\byear{2005}).
\btitle{Bayesian inference for origin-destination matrices of transport networks using the EM algorithm}.
\bjournal{Technometrics}
\bvolume{47}
\bpages{399--408}.
\bid{doi={10.1198/004017005000000283}, issn={0040-1706}, mr={2208309}}
\end{barticle}
%

\bptok{imsref}%
% NOT OUTPUTTED:
%   number = 4
%   doi = http://dx.doi.org/10.1198/004017005000000283
%   coden = TCMTA2
%   fjournal = Technometrics. A Journal of Statistics for the Physical, Chemical and Engineering Sciences
\endbibitem

%b23 ###
%b23 #&#
\bibitem[\protect\citeauthoryear{Liang and Yu}{2003}]{LiangYu03}
\begin{barticle}[author]
\bauthor{\bsnm{Liang},~\bfnm{G.}\binits{G.}} \AND
\bauthor{\bsnm{Yu},~\bfnm{B.}\binits{B.}}
(\byear{2003}).
\btitle{Maximum pseudo-likelihood estimation in network tomography}.
\bjournal{IEEE Trans. Signal Process.}
\bvolume{51}
\bpages{2043--2053}.
\end{barticle}
%

\bptok{imsref}%
\endbibitem

%b24 ###
%b24 #&#
\bibitem[\protect\citeauthoryear{Louis}{1982}]{Louis82}
\begin{barticle}[mr]
\bauthor{\bsnm{Louis},~\bfnm{Thomas~A.}\binits{T.~A.}}
(\byear{1982}).
\btitle{Finding the observed information matrix when using the EM algorithm}.
\bjournal{J. Roy. Statist. Soc. Ser. B}
\bvolume{44}
\bpages{226--233}.
\bid{issn={0035-9246}, mr={0676213}}
\end{barticle}
%

\bptok{imsref}%
% NOT OUTPUTTED:
%   url = http://links.jstor.org/sici?sici=0035-9246(1982)44:2<226:FTOIMW>2.0.CO;2-#&origin=MSN
%   number = 2
%   coden = JSTBAJ
%   fjournal = Journal of the Royal Statistical Society. Series B. Methodological
\endbibitem

%b25 ###
%b25 #&#
\bibitem[\protect\citeauthoryear{Maher}{1983}]{Maher83}
\begin{barticle}[mr]
\bauthor{\bsnm{Maher},~\bfnm{M.~J.}\binits{M.~J.}}
(\byear{1983}).
\btitle{Inferences on trip matrices from observations on link volumes: A {B}ayesian statistical approach}.
\bjournal{Transp. Res., Part B: Methodol.}
\bvolume{17}
\bpages{435--447}.
\bid{doi={10.1016/0191-2615(83)90030-9}, issn={0191-2615}, mr={0726928}}
\end{barticle}
%

\bptok{imsref}%
% NOT OUTPUTTED:
%   number = 6
%   doi = http://dx.doi.org/10.1016/0191-2615(83)90030-9
%   coden = TRBMDY
%   fjournal = Transportation Research. Part B. Methodological
\endbibitem

%b26 ###
%b26 #&#
\bibitem[\protect\citeauthoryear{Parry and Hazelton}{2013}]{ParryH13trb}
\begin{barticle}[author]
\bauthor{\bsnm{Parry},~\bfnm{K.}\binits{K.}} \AND
\bauthor{\bsnm{Hazelton},~\bfnm{M.~L.}\binits{M.~L.}}
(\byear{2013}).
\btitle{Bayesian inference for day-to-day dynamic traffic models}.
\bjournal{Transp. Res., Part B: Methodol.}
\bvolume{50}
\bpages{104--115}.
\end{barticle}
%

\bptok{imsref}%
\endbibitem

%b27 ###
%b27 #&#
\bibitem[\protect\citeauthoryear{{R Core Team}}{2013}]{citeR13}
\begin{bmisc}[author]
\bauthor{\bsnm{{R Core Team}}}
(\byear{2013}).
\bhowpublished{\textit{R: A Language and Environment for Statistical Computing}.
R Foundation for Statistical Computing,
Vienna, Austria.
ISBN 3-900051-07-0.}
\end{bmisc}
%

\bptok{imsref}%
\endbibitem

%b28 ###
%b28 #&#
\bibitem[\protect\citeauthoryear{Singhal and Michailidis}{2007}]{SinghalMichailidis07}
\begin{barticle}[mr]
\bauthor{\bsnm{Singhal},~\bfnm{Harsh}\binits{H.}} \AND
\bauthor{\bsnm{Michailidis},~\bfnm{George}\binits{G.}}
(\byear{2007}).
\btitle{Identifiability of flow distributions from link measurements with applications to computer networks}.
\bjournal{Inverse Problems}
\bvolume{23}
\bpages{1821--1849}.
\bid{doi={10.1088/0266-5611/23/5/004}, issn={0266-5611}, mr={2353317}}
\end{barticle}
%

\bptok{imsref}%
% NOT OUTPUTTED:
%   number = 5
%   doi = http://dx.doi.org/10.1088/0266-5611/23/5/004
%   coden = INPEEY
%   fjournal = Inverse Problems. An International Journal on the Theory and Practice of Inverse Problems, Inverse Methods and Computerized Inversion of Data
\endbibitem

%b29 ###
%b29 #&#
\bibitem[\protect\citeauthoryear{Tanner}{1996}]{Tanner96}
\begin{bbook}[mr]
\bauthor{\bsnm{Tanner},~\bfnm{Martin~A.}\binits{M.~A.}}
(\byear{1996}).
\btitle{Tools for Statistical Inference: Methods for the Exploration of Posterior Distributions and Likelihood Functions},
\bedition{3rd} ed.
\bpublisher{Springer},
\blocation{New York}.
\bid{doi={10.1007/978-1-4612-4024-2}, mr={1396311}}
\end{bbook}
%

\bptok{imsref}%
% NOT OUTPUTTED:
%   doi = http://dx.doi.org/10.1007/978-1-4612-4024-2
%   isbn = 0-387-94688-8
%   fpage = viii+207
\endbibitem

%b30 ###
%b30 #&#
\bibitem[\protect\citeauthoryear{Tebaldi and West}{1998}]{TebaldiWest98}
\begin{barticle}[mr]
\bauthor{\bsnm{Tebaldi},~\bfnm{Claudia}\binits{C.}} \AND
\bauthor{\bsnm{West},~\bfnm{Mike}\binits{M.}}
(\byear{1998}).
\btitle{Bayesian inference on network traffic using link count data}.
\bjournal{J.~Amer. Statist. Assoc.}
\bvolume{93}
\bpages{557--576}.
\bid{doi={10.2307/2670105}, issn={0162-1459}, mr={1631325}}
\bptnote{check related}%
\end{barticle}
%

\bptok{imsref}%
% NOT OUTPUTTED:
%   number = 442
%   doi = http://dx.doi.org/10.2307/2670105
%   coden = JSTNAL
%   fjournal = Journal of the American Statistical Association
\endbibitem

%b31 ###
%b31 #&#
\bibitem[\protect\citeauthoryear{Vanderbei and Iannone}{1994}]{VanderbeiIannone94}
\begin{bmisc}[author]
\bauthor{\bsnm{Vanderbei},~\bfnm{R.~J.}\binits{R.~J.}} \AND
\bauthor{\bsnm{Iannone},~\bfnm{J.}\binits{J.}}
(\byear{1994}).
\bhowpublished{An EM approach to OD matrix estimation.
Technical Report SOR 94-04, Princeton Univ., Princeton, NJ.}
\end{bmisc}
%

\bptok{imsref}%
% NOT OUTPUTTED:
%   publisher = Technical Report SOR 94-04, Princeton Univ.
\endbibitem

%b32 ###
%b32 #&#
\bibitem[\protect\citeauthoryear{Vardi}{1996}]{Vardi96}
\begin{barticle}[mr]
\bauthor{\bsnm{Vardi},~\bfnm{Y.}\binits{Y.}}
(\byear{1996}).
\btitle{Network tomography: Estimating source-destination traffic intensities from link data}.
\bjournal{J. Amer. Statist. Assoc.}
\bvolume{91}
\bpages{365--377}.
\bid{doi={10.2307/2291416}, issn={0162-1459}, mr={1394093}}
\end{barticle}
%

\bptok{imsref}%
% NOT OUTPUTTED:
%   number = 433
%   doi = http://dx.doi.org/10.2307/2291416
%   coden = JSTNAL
%   fjournal = Journal of the American Statistical Association
\endbibitem

%b33 ###
%b33 #&#
\bibitem[\protect\citeauthoryear{Veinott and Dantzig}{1968}]{VeinottDantzig68}
\begin{barticle}[mr]
\bauthor{\bsnm{Veinott},~\bfnm{Arthur~F.}\binits{A.~F.} \bsuffix{Jr.}} \AND
\bauthor{\bsnm{Dantzig},~\bfnm{George~B.}\binits{G.~B.}}
(\byear{1968}).
\btitle{Integral extreme points}.
\bjournal{SIAM Rev.}
\bvolume{10}
\bpages{371--372}.
\bid{issn={0036-1445}, mr={0232787}}
\end{barticle}
%

\bptok{imsref}%
% NOT OUTPUTTED:
%   fjournal = SIAM Review
\endbibitem

%b34 ###
%b34 #&#
\bibitem[\protect\citeauthoryear{Yai, Iwakura and Morichi}{1997}]{Yaietal97}
\begin{barticle}[author]
\bauthor{\bsnm{Yai},~\bfnm{T.}\binits{T.}},
\bauthor{\bsnm{Iwakura},~\bfnm{S.}\binits{S.}} \AND
\bauthor{\bsnm{Morichi},~\bfnm{S.}\binits{S.}}
(\byear{1997}).
\btitle{Multinomial probit with structured covariance for route choice behavior}.
\bjournal{Transp. Res., Part B: Methodol.}
\bvolume{31}
\bpages{195--207}.
\end{barticle}
%

\bptok{imsref}%
\endbibitem

%b35 ###
%b35 #&#
\bibitem[\protect\citeauthoryear{Ziegler}{1995}]{Ziegler95}
\begin{bbook}[mr]
\bauthor{\bsnm{Ziegler},~\bfnm{G{\"u}nter~M.}\binits{G.~M.}}
(\byear{1995}).
\btitle{Lectures on Polytopes}.
\bseries{Graduate Texts in Mathematics}
\bvolume{152}.
\bpublisher{Springer},
\blocation{New York}.
\bid{doi={10.1007/978-1-4613-8431-1}, mr={1311028}}
\end{bbook}
%

\bptok{imsref}%
% NOT OUTPUTTED:
%   doi = http://dx.doi.org/10.1007/978-1-4613-8431-1
%   isbn = 0-387-94365-X
%   fpage = x+370
\endbibitem

%b36 ###
%b36 #&#
\bibitem[\protect\citeauthoryear{Zuylen and Willumsen}{1980}]{VanZuylenWillumsen80}
\begin{barticle}[author]
\bauthor{\bsnm{Zuylen},~\bfnm{H.~J.~Van}\binits{H.~J.~V.}} \AND
\bauthor{\bsnm{Willumsen},~\bfnm{L.~G.}\binits{L.~G.}}
(\byear{1980}).
\btitle{The most likely trip matrix estimated from traffic counts}.
\bjournal{Transp. Res. Part B}
\bvolume{14}
\bpages{281--293}.
\end{barticle}
%

\bptok{imsref}%
\endbibitem
\end{thebibliography}
\end{document}